\documentclass[a4paper,11pt]{article}
\pdfoutput=1
\usepackage{jheppub}
\usepackage{datetime}
\usepackage{float}

\newcommand{\p}{{\partial}}
\newcommand{\zb}{{\bar z}}
\newcommand{\Zb}{{\bar Z}}

\newcommand{\pb}{{\bar \partial}}

\newcommand{\ep}{{\epsilon}}
\newcommand{\vep}{{\varepsilon}}
\newcommand{\rmd}{{\mathrm d}}
\newcommand{\w}{{\omega}}

\newcommand{\R}{{\mathbb{R}}}

\newcommand{\SL}{{SL}}
\newcommand{\SU}{{SU}}
\let\sl\relax
\newcommand{\sl}{{sl}}

\let\O\relax
\newcommand{\O}{{\cal O}}
\let\Re\relax
\DeclareMathOperator{\Re}{Re}

\DeclareMathOperator{\Sch}{Sch}
\DeclareMathOperator{\arcsinh}{arcsinh}

\DeclareMathOperator{\Tr}{Tr}
\DeclareMathOperator{\diag}{diag}
\DeclareMathOperator{\id}{id}

\DeclareMathOperator{\Diff}{Diff}

\title{AdS$_2$ Holography and Effective QFT}

\abstract{
We discuss AdS$_2$ quantum gravity from an unconventional perspective that emphasizes bulk geometry.
In our approach, AdS$_2$ has no boundary, there are no divergences that require renormalization, 
and the dilaton of JT-gravity can be omitted altogether. The result is the standard Schwarzian theory. However, it may be advantageous that our derivation just relies on conventional AdS/CFT correspondence and effective quantum field theory.
For example, it clarifies the symmetry 
breaking pattern. It also puts the non-compact AdS$_2$ 
topology on the same footing as compact Riemann surfaces.}

\author[a]{Sangmin Choi}
\author[b,c]{and Finn Larsen}

\emailAdd{sangmin.choi@polytechnique.edu}
\emailAdd{larsenf@umich.edu}
  
\affiliation[a]{CPHT, CNRS, \'Ecole Polytechnique, IP Paris, F-91128 Palaiseau, France.}
\affiliation[b]{Department of Physics and Leinweber Center for Theoretical Physics, \\University of Michigan, Ann Arbor, MI 48109-1120, USA.}

\affiliation[c]{Department of Physics and Stanford Institute for Theoretical Physics,\\Stanford University, Palo Alto, CA 94305, USA.}

\begin{document}


\maketitle


\section{Introduction}

The modern understanding of holographic correspondence between AdS$_2$ gravity and CFT$_1$ has matured to a level where most researchers adhere to a single prevailing narrative. 
On the boundary side, the theory emerges in the IR of a 1D quantum system, typically a version of SYK \cite{Sachdev:1992fk,Sachdev:2010um,Kitaev,Sachdev:2015efa,Jevicki:2016bwu,Jevicki:2016ito,Polchinski:2016xgd,Maldacena:2016hyu}. Meanwhile, the bulk theory is realized as JT gravity \cite{Jackiw:1982hg,Jackiw:1984je,Teitelboim:1983ux,Almheiri:2014cka,Maldacena:2016upp,Engelsoy:2016xyb,Sachdev:2019bjn}, perhaps coupled to conformal matter \cite{Engelsoy:2016xyb,Maldacena:2016upp}, and analyzed in conformal gauge. 
The purpose of this article is to study bulk AdS$_2$ quantum gravity from an unconventional point of view that highlights less developed aspects of the correspondence. For example, we never regularize AdS$_2$, or any fields therein, and we focus on bulk deformations, rather than the shape of the AdS$_2$ boundary. 

Our final results are not at all novel: the low energy dynamics of AdS$_2$, including its quantum corrections, is determined by the Schwarzian theory. However, in our considerations there is no gravitational action {\it a priori}, we do not from the outset presume the Jackiw-Teitelboim (JT) action \cite{Almheiri:2014cka,Maldacena:2016upp}, or anything else for that matter. Thus our starting point has no analogue of the dilaton that is a prominent ingredient of both JT gravity and the dimensional reduction of higher dimensional black holes. The corresponding feature is introduced much later, when we discuss matching with the UV theory in the sense of effective quantum field theory. Because of these and other differences, our approach gives a new perspective on some applications of AdS$_2$ quantum gravity that enter much current research. 

To make contact with standard considerations, consider the metric of a 2D spacetime in conformal gauge: 
\begin{equation}
ds^2 = e^{2\Phi(z, {\bar z})} dz d{\bar z}~. 
\label{eqn:confform}
\end{equation}
Under appropriate technical conditions, these metrics are representative of all 2D geometries, because the three components of a general 2D metric $g_{\mu\nu}$ can be taken to diagonal form by two diffeomorphisms $\xi^\mu$ acting on the two coordinates $x^\mu$. 
However, AdS$_2$ quantum gravity is qualitatively different from other 2D theories. A representative conformal factor is given by the conformal disc where
\begin{equation}
e^{2{\hat\Phi} (z, {\bar z})}   = \frac{4}{ (1 - z{\bar z})^2}~~~,~~~~|z|<1~. 
\label{eqn:confdisc} 
\end{equation} 
The divergence as $|z|\to 1$ necessitates the restriction to $|z|<1$. This in turn identifies the geometry as non-compact, 
the ``boundary" at $|z|=1$ is not part of the geometry.  

The standard approach to non-compactness of AdS$_2$ is to regulate the conformal disc \eqref{eqn:confdisc} by introducing a boundary regulator at $|z|<1$ \cite{Maldacena:2016upp,Kitaev:2017awl}. The regulated geometry is compact, with the topology of a disc. The length of the boundary diverges as $|z|\to 1$, but low energy correlation functions depend only on the ``shape" of this artificial boundary, or more precisely its shape modulo the $\SL(2)$ isometry group of AdS$_2$. The upshot is that, following the renormalization program introduced in any textbook on quantum field theory, there is a low energy theory that depends only on the shape of the boundary.

Our complementary strategy addresses the non-compact geometry explicitly. 
We do not encounter divergences at intermediate stages so no regularization is needed and, therefore, we can maintain the AdS$_2$ topology throughout. The central goal is to focus on the configuration space of AdS$_2$ geometries, rather than its boundary. This approach leads to considerations that are more commonly associated with critical string theory, especially its implementation on world-sheets with genus $g\geq 2$.

Following the critical string theory template \cite{Green:1987sp,Lust:1989tj,Polchinski:1998rq,Becker:2006dvp}, an important question is whether, after using two variables' worth of diffeomorphisms to attain the conformal gauge \eqref{eqn:confform} with conformal factor \eqref{eqn:confdisc}, any residual diffeomorphisms remain. Such ``unneeded" reparametrizations are generated by the Conformal Killing Vectors (CKV's) of the 2D geometry. Locally, AdS$_2$ is 
reminiscent of the two-sphere $S^2$, in that their isometries generate three-dimensional algebras $\SL(2)$ and $\SU(2)$, respectively, and locally their CKV's both generate the obvious complexifications. The key distinction is global: on the two-sphere $S^2$ the CKV's are legitimate, because they are normalizable, but AdS$_2$ is non-compact and the conformal factor \eqref{eqn:confdisc} diverges in a manner that invalidates all candidate CKV's. In short, according to the principles of our approach, there are {\it no} Conformal Killing Vectors on AdS$_2$, and certainly no Killing Vectors . 

Continuing with critical string world-sheet theory as inspiration, a second important question is whether two variables' worth of diffeomorphisms is truly sufficient to reach the conformal gauge \eqref{eqn:confform}. Locally, this is unquestionable, solutions to Beltrami's differential equations can be presented explicitly in terms of certain integrals. However, on compact Riemann surfaces with genus $g\geq 2$, it is well-known that some candidate diffeomorphisms are formal, they are not globally well-defined. Such ``illegitimate" diffeomorphisms are not mere changes of coordinates, they are generators of physically distinct geometries. The configuration space obtained this way, usually referred to as moduli space, has $6g-6$ real dimensions and interesting global features. 

We analogously inquire whether all AdS$_2$-geometries of the form
\begin{equation}
ds^2_2 =  \frac{4}{ (1 - z{\bar z})^2}dz d{\bar z} + (\delta g_{zz}(z, {\bar z})dz^2 + {\rm c.c.}) ~,
\end{equation} 
can be transformed to the fiducial case where $\delta g_{zz}=0$ by some diffeomorphism. We show that this is {\it not} possible for 
\begin{equation}
ds^2_2 =  \frac{4}{ (1 - z{\bar z})^2}dz d{\bar z} + ( h (z) dz^2 + {\rm c.c.}) 
\label{eqn:holoproto} 
\end{equation} 
where $h(z)$ is an arbitrary infinitesimal {\it holomorphic} function, but otherwise it is. Importantly, the precise claim is that {\it formally} diffeomorphisms that remove the functions $h(z)$ can be constructed, but such diffeomorphisms are not normalizable because of the divergent conformal factor \eqref{eqn:confform}. Therefore, metrics with distinct holomorphic function $h(z)$ cannot be identified with one another, and so they parameterize the physical configuration space of AdS$_2$ geometries. Comparing with the moduli space of geometries with genus $g\geq 2$, the configuration space of AdS$_2$ geometries arrived at this way is a $g\to\infty$ limit, a space known as 
the {\it universal} Teichm\"{u}ller space.

One of the exciting applications of JT gravity is the augmentation of the conventional topological classification of compact 2D Riemann surfaces, involving handles, (compact) boundaries, and fixed points, to include boundaries that are {\it asymptotically} AdS$_2$ \cite{Saad:2019lba,Stanford:2019vob,Penington:2019kki,Kapec:2019ecr,Saad:2022kfe,Stanford:2022fdt,Iliesiu:2019lfc,Goel:2020yxl,Hsin:2020mfa,Iliesiu:2021ari,Blommaert:2021fob,Iliesiu:2021are,Iliesiu:2022kny,Maxfield:2020ale,Mertens:2020hbs,Turiaci:2020fjj,Rosso:2021orf,Mertens:2022irh}. In the literature, these non-compact ``trumpet" boundaries are associated with Schwarzian actions that classify the boundary shape ${\rm Diff}(S^1)/\SL(2)$. Our non-standard presentation
focuses on the freedom in the 2D geometry which, as indicated in 
\eqref{eqn:holoproto} amounts to a holomorphic function. This point of view
puts AdS$_2$ boundaries on the same footing as other topologies, 
the entire physical configuration space is parametrized by quadratic holomorphic differentials. 
This perspective complements the narrative that stresses the ``shapes" of the boundary \cite{Maldacena:2016upp,Engelsoy:2016xyb,Kitaev:2017awl,Sarosi:2017ykf,Saad:2018bqo,Saad:2019lba,Stanford:2019vob}.

In recent discussions of AdS$_2$ quantum gravity, it is usually taken for granted that the Schwarzian effective theory {\it must} descend from JT-gravity, or at least some gravitational action that has an Einstein term and an effective dilaton with a slope that introduces a scale. Our approach is consistent with this lore, in that the Schwarzian action {\it may} have come about this way. However, we derive equivalent results using {\it only} the symmetries of AdS$_2$ geometry and the well-established logic of standard effective QFT. This may have some conceptual advantages.

To introduce how this is possible, note that the metric \eqref{eqn:holoproto} appears to be particularly simple when $h(z)=0$.
However, this impression is an artifact of the coordinates we employ: configurations with different functions $h(z)$ are physically distinct, because they are related by non-normalizable diffeomorphisms, but they are all equivalent in that none of them is preferred over any other.  For example, we can assign the same energy to all these geometries, and the spectrum of fluctuations is the same around any one of them. This is the set-up for spontaneously broken symmetry. 

As usual, the symmetry breaking pattern identifies the entire low energy effective QFT. The modes that dominate at low energy are identified by Goldstone's theorem, and interpreted as slow oscillations in the space of degenerate vacuum configurations parametrized by the holomorphic function $h(z)$. Non-linear realization of symmetry determines the interactions between these modes in a manner that is reviewed pedagogically in many texts, including
\cite{Manohar:1996cq, Rothstein:2003mp, Burgess:2007pt}.
In our presentation we seek to follow the standard examples of effective QFT as close as possible, to highlight that the emergence of a Schwarzian description is universal and not tied to a gravitational action.

Any effective QFT depends on a dimensionful scale that serves as a coupling for a derivative expansion.
For the Schwarzian effective theory we can take this parameter as $C_T/T$. The nomenclature refers to high temperature, where the scale is the heat capacity in units of the temperature, but the same parameter also governs the low temperature limit where, in general, the dynamics is dramatically different
 \cite{Stanford:2017thb,Charles:2019tiu,Iliesiu:2020qvm,Boruch:2022tno}. The dimensionful coupling constant is arbitrary, according to principles intrinsic to the IR theory. It constitutes a single piece of data inherited from the UV theory at slightly larger energy and can be determined only by ``matching", typically by comparing a physical observable that is convenient to compute in both the effective theory and its UV progenitor. The power of effective QFT is that non-linearly realized symmetry demands that the same effective parameter sets the scale of any other process. 

As we have discussed, the subject of this article was pursued by numerous authors over the last few years, from many different points of view. Our goal is to introduce perspectives from 2D bulk geometry that we think are important, yet not emphasized in recent discussions. In view of the relative maturity of the subject we have divided the article in four sections that have some degree of independence: 

\begin{itemize}
\item
In section \ref{sec:AdS_2} we discuss the AdS$_2$ geometry using the {\it Fefferman-Graham} expansion. This gives a version of the modern AdS$_2$/CFT$_1$ correspondence that adheres closely to the template that is familiar from AdS$_{d+1}$/CFT$_d$ correspondence with $d>1$.
For example, the CFT$_1$ ``lives" on a conformal boundary \cite{Maldacena:1997re,Witten:1998qj}. We emphasize issues that 
are novel when $d=1$. 

\item
In section \ref{sec:worldsheet} we develop the AdS$_2$ geometry in
{\it holomorphic} coordinates. The principle is much the same 
as in section \ref{sec:AdS_2} but, as discussed  
after \eqref{eqn:holoproto}, the technical implementation follows ideas that are usually associated with the string theory world-sheet. 
Despite the apparent differences, it turns out that the results in
sections \ref{sec:AdS_2} and \ref{sec:worldsheet} are gauge equivalent.
We show that by constructing an explicit coordinate transformation 
that relates them and is {\it normalizable}, and therefore a genuine symmetry. 

\item
In section \ref{sec:gaugethy} we study the first order formalism for gravity and its equivalent gauge theory formulation. As in previous sections, we do not introduce a cut-off, opting instead for using the non-compact geometry to distinguish between symmetries
and solution generating gauge transformations.
The gauge theory approach gives a derivation of the Schwarzian action that is particularly satisfying, in that it follows ad verbatim from the non-linear sigma-model. 

\item
Finally, in section \ref{sec:AdS3toAdS2} we revisit AdS$_2$ quantum gravity as a dimensional reduction of AdS$_3$. The interpretation of AdS$_2$ quantum gravity as a slice of AdS$_3$ identifies the dual CFT$_1$ as a chiral sector of a CFT$_2$. As in previous sections our emphasis is on normalizability and, at this point, also on the consistency between various approaches. 
\end{itemize}

In view of this rather lengthy introduction, we do not include a final discussion.

\section{Distinct AdS\texorpdfstring{$_2$}{} Geometries and the AdS/CFT Correspondence}
\label{sec:AdS_2}

In this section we discuss the space of physically distinct AdS$_2$ geometries by following the strategy that is familiar from the AdS/CFT correspondence in higher dimensions as closely as possible.
We highlight how geometries that are formally related to one another by diffeomorphisms are in fact physically distinct. That is possible because
AdS$_2$ is not compact.

\subsection{AdS\texorpdfstring{$_2$}{} in Fefferman-Graham Gauge} 

We present the baseline Euclidean AdS$_2$ metric as
\begin{align}
	ds^2 = d\rho^2 + \sinh^2\rho\,d\tau^2
	\label{eqn:ads2_sinh2rho}
	\ .
\end{align}
Here $\rho$ is a radial coordinate pointing outwards and $\tau$ parametrizes the angle on the concentric circles of constant $\rho$. Near the origin $\rho=0$ regularity imposes the periodicity $\tau\sim \tau+2\pi$. The geometry in this region is similar to the origin of a disc with standard radial coordinates. 

The geometries of AdS$_2$ and a disc differ for larger $\rho$,
because the former has constant negative curvature, and the latter is flat. 
After regularization by a cut-off at very large $\rho$, AdS$_2$ and the disc become equivalent topologically, and geometrically they just differ by a conformal factor. However, in the context of AdS$_2$, the introduction of a cutoff is not an inconsequential technical device: it changes the topology of the geometry, and the conformal structure diverges as the cut-off is removed. In our discussion below we study AdS$_2$ without introducing a cut-off. 

In keeping with the standard AdS/CFT procedure, we seek other AdS$_2$ metrics that deform \eqref{eqn:ads2_sinh2rho} within a gauge that is analogous to the Fefferman-Graham gauge in higher dimensions:
\begin{align}
	ds^2 = d\rho^2 + g_{\tau\tau}(\tau,\rho) d\tau^2
	\ ,
	\label{eqn:fggauge}
\end{align}
where
\begin{align}
g_{\tau\tau}(\tau,\rho)
= \frac{1}{4} e^{2\rho} + g_0(\tau)  + g_1 (\tau) e^{-2\rho} + \cdots. 
	\label{eqn:gttgauge}
\end{align}
Thus the Fefferman-Graham gauge fixes the components $g_{\rho\rho}=1$ and $g_{\tau\rho}=0$. The remaining variable $g_{\tau\tau}$ depends on both coordinates, with a perturbative expansion at large $\rho$ designed to preserve the asymptotic behavior near the conformal boundary at $\rho\to\infty$, with the precise fall-off conditions discussed shortly. 
We henceforth refer to this entire structure as the Fefferman-Graham gauge.

Coordinate transformations that preserve the Fefferman-Graham gauge may be residual gauge symmetries (unphysical redundancies), physical symmetries (relations between different geometries), or 
formal maps between geometries that do not belong to the same physical configuration space. To illustrate these important distinctions, we consider a reparametrization of the Euclidean time $\tau$ to some function $f(\tau)$. Since the initial coordinate is periodic $\tau\sim \tau+2\pi$, the function 
$f$ must be $2\pi$-periodic in $\tau$ to be well-defined. 
Also, for the new ``time" $f$ to be periodic as well, 
$f$ should itself 
be valued on $S^1$ rather than $\R$.
Finally, regularity of the curve on each concentric circle
demands that the ``velocity'' (the first derivative of $f$) vanishes nowhere. Thus, we impose the following conditions
\begin{align}
	f(\tau) = f(\tau + 2\pi)
	\ ,\qquad
	f(\tau)\equiv f(\tau)+2\pi
	\ ,\qquad f'(\tau)>0
	\ ,
	\label{eqn:f_conditions}
\end{align}
on $f$ to ensure that these reparametrizations are well-behaved.

Even with these conditions imposed, a simple coordinate transformation that just replaces $\tau$ with $f(\tau)$ is not acceptable, as it modifies the metric $\delta g_{\tau\tau} \sim \O(e^{2\rho})$ and so violates the leading order of the gauge condition \eqref{eqn:gttgauge}. This condition cannot be relaxed because 
a physical deformation of the metric must be normalizable in the sense that 
\begin{equation}
    \int d^2x\sqrt g ~|\delta g|^2<\infty~. 
    \label{eqn:delgnorm}
\end{equation}
Non-normalizable departures from the baseline metric \eqref{eqn:ads2_sinh2rho}
correspond to geometries that do not belong to the same classical phase space. The condition 
\eqref{eqn:delgnorm}
means 
$\delta g_{\tau\tau}$ can be at most of order $\O(1)$ at large $\rho$. 
This explains why the leading order $\O(e^{2\rho})$ in the gauge condition \eqref{eqn:gttgauge} is constant, it cannot involve a function of $\tau$.

These deficiencies can be addressed by not only transforming $\tau$, but also changing the radial coordinate $\rho$ so $\sinh^2(\rho)\to f'(\tau)^{-2}\sinh^2(\rho)$. This cancels the leading correction to $\delta g_{\tau\tau}$ and renders the residual metric deformation $\delta g_{\tau\tau}\sim \O(1)$ normalizable.

Unfortunately, the remedy that cures the asymptotic behavior of $g_{\tau\tau}$ modifies $g_{\rho\rho}$ and $g_{\tau\rho}$,  in a manner that violates the gauge conditions. This problem does
not affect the asymptotic behavior at large $\rho$, and the gauge can be restored by adding subleading corrections to $\tau\to f(\tau) + \cdots$. This in turn necessitates corrections to $\sinh^2(\rho)\to f'(\tau)^{-2}\sinh^2(\rho)(1+\cdots)$, and so on and so forth. Continuing in this manner, we obtain the permissible residual coordinate transformations as a perturbative series in
$e^{-2\rho}$:
\begin{align}
	\tau
	\quad\to\quad&
		f(\tau)
		- 2f''(\tau) e^{-2\rho}
		- 2f''(\tau)\left(f'(\tau)-\frac{f''(\tau)^2}{f'(\tau)^2}\right)e^{-4\rho}
		+ \cdots
	\label{series_tau}
	\ ,\\
	\sinh^2(\rho)
	\quad\to\quad&
		\frac{\sinh^2(\rho)}{f'(\tau)^2}\left(
			1
			+ 2\left(
				1
				- f'(\tau)^2
				+ \frac{f''(\tau)^2}{f'(\tau)^2}
			\right)e^{-2\rho}
			+ \cdots
		\right)
	\label{series_rho}
	\ .
\end{align}
Continuing to a few more orders, we find the large-$\rho$ expansion of the coordinate transformations (see also \cite{Castro:2019vog}):
\begin{align}
\begin{split}
	\tau
	\quad\to\quad&
		f(\tau)
		- \arctan\left[
			\frac{2e^{-2\rho}f''(\tau)}{1-e^{-2\rho}\left(f'(\tau)^2-\frac{f''(\tau)^2}{f'(\tau)^2}\right)}
		\right]
	\ ,\\
	\rho
	\quad\to\quad&
		\arcsinh\left[
			\frac{e^{\rho}}{2f'(\tau)}
			\sqrt{
				\left(
					1
					+ e^{-2\rho}\left(
						f'(\tau)^2 + \frac{f''(\tau)^2}{f'(\tau)^2}
					\right)
				\right)^2
				- 4e^{-2\rho} f'(\tau)^2
			}
		\right]
	\ .
	\label{2d_finite_real}
\end{split}
\end{align}
Indeed, these exact transformations preserve the Fefferman-Graham gauge. They transform the metric \eqref{eqn:ads2_sinh2rho} to
\begin{align}
	ds^2
	& = 
		d\rho^2
		+ \frac{1}{4}\left[
			e^\rho
			- 2e^{-\rho}\left(
				\Sch(f,\tau)
				+ \frac{1}{2}f'(\tau)^2
			\right)
		\right]^2 d\tau^2
	\ ,
	\label{ds2_SchTan}
\end{align}
where $\Sch(\cdot,\cdot)$ denotes the Schwarzian derivative,
\begin{align}
	\Sch(f,\tau) = \frac{f'''(\tau)}{f'(\tau)} - \frac{3f''(\tau)^2}{2f'(\tau)^2}
	\ .
	\label{eqn:Sch}
\end{align}
The computation shows that this line element is exactly AdS$_2$ for any function $f(\tau)$.
We note for future reference that, for an infinitesimal deviation $\vep(\tau)$ away from the identity, we can write $f(\tau) = \tau + \vep(\tau)$ and the metric \eqref{ds2_SchTan} becomes
\begin{align}
    ds^2 = d\rho^2 + (\sinh\rho)^2 d\tau^2 -2 e^{-\rho}\sinh\rho \left(\vep'(\tau) + \vep'''(\tau)\right) d\tau^2
    + \O(\vep^2)
    \label{ds2_SchTan_infinitesimal}
    \ .
\end{align}
to linear order in $\vep(\tau)$.

To summarize, in Fefferman-Graham gauge there is a 
family of Euclidean AdS$_2$ geometries
\eqref{ds2_SchTan}. They are generated by the diffeomorphisms \eqref{2d_finite_real} acting on the two coordinates $(\tau,\rho)$. Thus the family is 
parametrized by a single function $f(\tau)$ that satisfies the conditions \eqref{eqn:f_conditions}. It can be interpreted as a reparametrization of the unit circle at asymptotic infinity.

\subsection{Large Diffeomorphisms}

All coordinate transformations change the precise form of the metric but usually such
deformations are unphysical because they leave the underlying geometry invariant. If the coordinate transformations \eqref{2d_finite_real} were gauge symmetries in this sense, the AdS$_2$ metrics \eqref{ds2_SchTan} with distinct $f(\tau)$ would be identified and correspond to a single point in the physical configuration space. 
In this subsection we show that, to the contrary, they are ``large" diffeomorphisms: 
the coordinate transformations \eqref{2d_finite_real} are generated by non-normalizable vector fields. It follows that \eqref{ds2_SchTan} is a family of physically distinct metrics.

An infinitesimal diffeomorphism $\delta x^\mu$ is normalizable if 
\begin{equation}
    \int d^2x\sqrt g ~|\delta x|^2 < \infty~. 
     \label{eqn:delxnorm}
\end{equation}
This condition is stronger than its analogue  \eqref{eqn:delgnorm} for the deformation of the metric because the norm-squared $|\delta x|^2 = g_{\mu\nu}\delta x^\mu \delta x^\nu$ involves a single factor of the baseline metric \eqref{eqn:ads2_sinh2rho} that diverges as the asymptotic boundary $\rho\to\infty$, while the norm 
$|\delta g|^2$ requires two such factors. Large diffeomorphisms that violate \eqref{eqn:delxnorm} but
induce changes in the metric that are normalizable according to \eqref{eqn:delgnorm} generate motions in the physical configuration space. 

The finite diffeomorphism \eqref{2d_finite_real} depends on a function $f$ that describes reparametrization as $\tau \to f(\tau)$
for asymptotically large $\rho$. The nearly trivial reparametrization is $f(\tau) = \tau + \vep(\tau)$ with infinitesimal $\vep$. Inserting it in  \eqref{2d_finite_real} and expanding to linear order in $\vep$ for any $\rho$ yields: 
\begin{align}
	\delta\tau = \vep(\tau) + \frac{2\vep''(\tau)}{1-e^{2\rho}}
	\ ,\qquad
	\delta\rho = -\vep'(\tau)
	\ .
	\label{diff_v}
\end{align}
The resulting norm-squared of $\delta x^\mu = (\delta\tau, \delta\rho)$ is not normalizable:
\begin{align}
	\int d^2x \sqrt g |\delta x|^2
	\ &\sim\ 
		\int d\tau d\rho\, \vep(\tau)^2 \sinh^3\rho
	\to \infty
	\ ,
\end{align}
and therefore the finite diffeomorphism \eqref{2d_finite_real} is non-normalizable as well. On the other hand, \eqref{ds2_SchTan} gives the metric deformation $\delta g_{\tau\tau} \sim -(\vep'+\vep''')$
generated by this reparametrization, and it  
is normalizable:
\begin{align}
	\int d^2x \sqrt g |\delta g_{\tau\tau}|^2
	\ &\sim\ 
		\int d\tau d\rho\, (\vep'+\vep''')^2\sinh^{-3}\rho
	< \infty
	\ . 
\end{align}
Thus the entire metric family \eqref{ds2_SchTan} is physical even though its elements are generated by the reparametrizations \eqref{2d_finite_real}, because the latter are not normalizable.

\subsection{Topology: the AdS\texorpdfstring{$_2$}{} Wormhole}
\label{subsec:wormhole}
The metrics \eqref{ds2_SchTan} realize a large family of Euclidean AdS$_2$ geometries, 
as many as there are functions $f(\tau)$ satisfying \eqref{eqn:f_conditions}. However, the family is not exhaustive, there exists AdS$_2$ geometries that do not belong to it. As an example, we consider
\begin{align}
	ds^2
	&=	d\rho^2 + \cosh^2\rho\, d\tau^2
	~,
	\label{eqn:wormh}
\end{align}
which indeed has $g_{\rho\rho}=1$, $g_{\rho\tau}=0$, and the asymptotic behavior \eqref{eqn:gttgauge}. This Euclidean AdS$_2$ geometry is interesting, because it features an AdS$_2$ boundary at both $\rho\to\infty$ and $\rho\to-\infty$. Thus it has wormhole topology.

This example motivates studying global features of the explicit AdS$_2$ geometries \eqref{ds2_SchTan}. This is not straightforward, because the
baseline geometry \eqref{eqn:ads2_sinh2rho}
is ill-defined at the origin of AdS$_2$ where $\rho=0$. The obstacle is just the familiar one facing polar coordinates; the $\tau$ coordinate can take any value $\tau\in [0,2\pi[$ at $\rho=0$, but for a point it should take a single value. Therefore, the Fefferman-Graham coordinates obscure the map of the neighborhood near the origin under the large diffeomorphism \eqref{2d_finite_real}. 

In some cases we can do better by introducing the function
\begin{align}
	F(\tau) = \tan\frac{f(\tau)}{2}
	\ .
	\label{eqn:Fdefi}
\end{align}
The identity of Schwarzian derivatives,
\begin{align}
	\Sch\left(\tan\frac{f(\tau)}{2},\tau\right)
	&=
	\Sch(f,\tau)
	+ \frac{1}{2}f'(\tau)^2
	\ ,
\end{align}
simplifies the expression \eqref{ds2_SchTan} of the metric such that its dependence on a function is given in terms of a single Schwarzian derivative,
\begin{align}
	ds^2
	=
		d\rho^2
		+ \frac{1}{4}\Big(
			e^\rho
			-2e^{-\rho}\Sch(F,\tau)
		\Big)^2d\tau^2
	\ .
	\label{eqn:Fmetrc}
\end{align}
This metric is equivalent to \eqref{ds2_SchTan}, as long as $F(\tau)$ complies with the analogues of the conditions \eqref{eqn:f_conditions}.
The last two conditions in \eqref{eqn:f_conditions} translate to the following two conditions on $F(\tau)$:
\begin{align}
	F(\tau) = F(\tau + 2\pi)
	\ ,\qquad
	F'(\tau)>0
	\ ,
	\label{eqn:Fpeiod}
\end{align}
where in the second inequality we used $f'(\tau) = \frac{2F'(\tau)}{1+F(\tau)^2}$.
To translate the first condition of \eqref{eqn:f_conditions}, note that the image of $f(\tau)$ is the entire interval $[0,2\pi[$. Therefore, since the tangent function maps the interval $(0,\pi)$ to $(-\infty,\infty)$,
the function $F(\tau)$ 
\eqref{eqn:Fdefi}
traverses the entire real line $\R$, rather than the circle $S^1$. Moreover, 
because the original image was $S^1$, implemented by the periodicity $f(\tau) = f(\tau)+2\pi$, the two infinities $-\infty$ and $\infty$ must be identified as the same geometrical point. Evidently, the map \eqref{eqn:Fpeiod} yields new coordinates
that are inadequate at loci where $f(\tau)=\pi\mod 2\pi$. This is unsurprising, reparametrization cannot by itself resolve a singular coordinate system. 

The benefit added by the parametrization in terms of $F(\tau)$ is generality. 
Periodicity of the Euclidean time $\tau$ can be interpreted as thermal boundary conditions. It is preserved by the family of diffeomorphisms \eqref{2d_finite_real} parametrized by $f(\tau)$. The description in terms of $F(\tau)$ is equivalent if the analogous periodicity condition, spelled out in the preceding paragraph, is satisfied. However, we may consider more general $F(\tau)$ that obey relaxed boundary conditions. Such geometries remain AdS$_2$ locally, but they are not equivalent to the baseline metric 
\eqref{eqn:ads2_sinh2rho}. 

Our motivating example, the wormhole geometry \eqref{eqn:wormh}, is indeed reproduced by 
the transformed metric \eqref{eqn:Fmetrc} with the map $F(\tau)=e^\tau$ that yields $\Sch(F(\tau),\tau)=-\frac{1}{2}$. Importantly, this $F(\tau)$ does not satisfy the periodicity condition in \eqref{eqn:Fpeiod}, nor is its image the entire real line. Therefore, the wormhole geometry is not related to the baseline metric \eqref{eqn:ads2_sinh2rho} by a legitimate coordinate transformation. 

As the wormhole example shows, a coordinate transformation that violates 
boundary conditions is a map to an inequivalent geometry. 
However, the map may nevertheless carry the large diffeomorphisms \eqref{2d_finite_real} that generate distinct geometries from the baseline metric \eqref{eqn:ads2_sinh2rho} into generators of an analogous family for the inequivalent target geometry. That is because the normalization conditions 
(\ref{eqn:delgnorm}, \ref{eqn:delxnorm}) that make diffeomorphisms ``large", and so generators of a physical configuration space are preserved by such transformations. In the next section we will 
exploit this strategy in a particular case.

In summary, Fefferman-Graham coordinates exhibit the geometry near the asymptotic boundary explicitly and conveniently but their continuation 
to the origin of AdS$_2$ is ambiguous. 
In the next section we discuss global aspects of AdS$_2$ geometries using another approach.

\section{AdS\texorpdfstring{$_2$}{} Geometry as a World-sheet}\label{sec:worldsheet}

In this section we reconsider physically distinct AdS$_2$ geometries following the strategy that is familiar from world-sheet string theory. We employ complex coordinates and inquire whether conformal gauge fixes the gauge entirely, and to what extent it can be reached in the first place. 

\subsection{Conformal Gauge and Non-compactness of AdS\texorpdfstring{$_2$}{}}
\label{subsec:confgauge}
The baseline AdS$_2$ metric \eqref{eqn:ads2_sinh2rho} can be recast as
\begin{align}
    ds^2 = \frac{4dzd\zb}{(1-z\zb)^2}~,
    \label{ads2_compl}
\end{align}
by introducing holomorphic coordinates 
\begin{align}
z=e^{i\tau}\tanh\frac{\rho}{2}~.
 \label{ztaurho}
\end{align}
This is advantageous for some purposes. For example, the Cartesian coordinates $(z,{\bar z})$ can be continued to the origin $z = {\bar z}=0$ where the geometry becomes manifestly regular but the radial coordinates $(\rho,\tau)$ break down. 

The metric \eqref{ads2_compl} is in {\it conformal gauge}: it is the flat 2D geometry 
$ds^2= 4dz d{\bar z}$ multiplied by a conformal factor 
$e^{2\Phi}=(1-z\zb)^{-2}$ that renders the curvature negative and constant. 
This form of the metric is reminiscent of world-sheet string theory. For example, the superficially similar conformal factor $e^{2\Phi} = (1+z\zb)^{-2}$ corresponds to the sphere $S^2$. 
The feature that is qualitatively different from familiar world-sheet considerations is 
the non-compactness of AdS$_2$. 

It is instructive to work out symmetries explicitly. The ``Killing vectors" for AdS$_2$ \eqref{ads2_compl} are 
\begin{equation}
K^{\bar z}_a =  - i ( 1,{\bar z}, {\bar z}^2)~~~~,~~K^z_a = i(z^2, z, 1)~,~~~
a=1, 0, -1 ~.
\label{eqn:KVs}
\end{equation}
They formally satisfy the Killing equations
\begin{equation}
\partial_\mu K_{a\nu} + \partial_{\nu} K_{a\mu}  = 0~,~~\mu, \nu = z, {\bar z}~.
\label{eqn:KVeqns}
\end{equation}
For $\mu=z, \nu={\bar z}$ (and {\it vice versa}) this follows from the obvious holomorphic structure of \eqref{eqn:KVs}. This shows that diffeomorphisms generated by \eqref{eqn:KVs} respect the conformal factor in \eqref{ads2_compl}.
To verify the fully holomorphic components (and their complex conjugates), the nontrivial Christoffel symbols are
$$
\Gamma^z_{zz} = g^{z{\bar z}} \partial_z g_{z{\bar z}} = \frac{2 {\bar z}}{1-|z|^2}~,
$$
and its complex conjugate. This gives the covariant derivatives 
$$
\nabla_z K_{az} = ( \partial_z - \Gamma^z_{zz})K_{az} 
=  - i \left( \partial_z - \frac{2 {\bar z}}{1-|z|^2}\right) \frac{2(1, {\bar z}, {\bar z}^2)}{(1 - |z|^2)^2} = 0 ~,
\label{eqn:KVsD}
$$
as claimed.
This aspect of \eqref{eqn:KVeqns} shows that diffeomorphisms generated by \eqref{eqn:KVs} preserve the gauge condition that \eqref{ads2_compl} has no fully holomorphic (or fully anti-holomorphic) component.

However, the Killing vectors \eqref{eqn:KVs} are formal, they do
not generate symmetries even though they satisfy the Killing equations. That is because they are {\it not normalizable} in the metric \eqref{ads2_compl}, the metric components $g_{z{\bar z}}\to\infty$ as $|z|\to 1$. This situation differs from the analogous computation on a sphere $S^2$ that is discussed in many textbooks \cite{Polchinski:1998rq,Lust:1989tj}. It also differs from the standard analysis after introducing a cut-off $|z|<1$ on the geometry \cite{Maldacena:2016upp}, because such a cut-off renders the Killing vectors normalizable. We analyse AdS$_2$ as a geometry in its own right, and for that it is essential that we allow for the entire non-compact space $|z|<1$. From this point of view there are {\it no} Killing vectors in AdS$_2$. 

The {\it conformal} Killing vectors (CKV's) are closely related to the Killing 
vectors, but their physical significance is quite different\footnote{Strictly speaking, Killing vectors are special cases of Conformal Killing vectors where the conformal factor is trivial. When we imply that Killing vectors and CKV's are non-overlapping classes, we have in mind the CKV's that are not also Killing vectors.}
They have 
components\footnote{We pick orientation so $\epsilon^{z{\bar z}}  = \frac{i}{\sqrt{g}}$ }
\begin{equation}
C_{az} = - i K_{az}~~,~~ C_{a{\bar z}} =  i K_{a{\bar z}}~.
\label{eqn:CvsK}
\end{equation}
The formulae worked out for the Killing vectors immediately give
$$
\nabla_z C_{az} =\nabla_{\bar z} C_{a{\bar z}}=0~.
$$
Diffeomorphisms generated by CKV's therefore preserve the part of the conformal gauge condition that posits a purely off-diagonal metric in holomorphic coordinates. However, in contrast to the Killing vectors \eqref{eqn:KVs}, diffeomorphisms generated by the CKV's 
\eqref{eqn:CvsK} do deform the diagonal part of the metric 
\begin{align}
\delta g_{z{\bar z}}= \partial_z C_{a{\bar z}} + \partial_{\bar z} C_{a z}  &= 
\frac{2(2z, 1, 0 ) + 2(0, 1, 2{\bar z} )}{(1 - |z|^2)^2} + 
\frac{4(z^2, z, 1){\bar z}  + 4(1, {\bar z}, {\bar z}^2)z}{(1 - |z|^2)^3}~,\cr
&= \frac{4(2z, 1+z{\bar z}, 2{\bar z} )}{(1 - |z|^2)^3} = 2g_{z{\bar z}}X_a~,
\label{eqn:CKVgdef}
\end{align}
where the potentials $X_a$ generating the flows of the vector fields satisfy
\begin{eqnarray}
g^{z{\bar z}} \partial_{\bar z} X_1 &=&  1~,
 \cr
g^{z{\bar z}}\partial_{\bar z} X_0 &=& z ~,
\cr
g^{z{\bar z}}\partial_{\bar z} X_{-1} &=&  z^2~.
\end{eqnarray}
Thus the components of the vector field $g^{{\bar z}z}\partial_{\bar z}X_a$ are the pure powers $z^n$ with $n=0, 1, 2$. The
potentials $X_a$
satisfy: 
\begin{align}
    \left( \nabla_z \nabla_{\bar z} + \nabla_{\bar z}  \nabla_z  - 2 g_{z{\bar z}} \right) X_a = 0 ~. 
\label{eqn:Xeqn}
\end{align}
In this expression all covariant derivatives $\nabla_\mu$ act as ordinary derivatives $\partial_\mu$. The equation identifies the $X_a$ as free scalar fields with $m^2=2$. 

The Conformal Killing vectors $(C^z_a, C^{\bar z}_a)$ are non-normalizable, just like the Killing vectors $(K^z_a, K^{\bar z}_a)$. Indeed, they differ only by the relative sign between holomorphic and anti-holomorphic components, and this makes no difference for normalizability. From our point of view we therefore reject both as unphysical. Thus there are {\it no residual diffeomorphism symmetries} after gauge fixing AdS$_2$ to conformal gauge \eqref{ads2_compl}.

Even though our principles lead us to reject both Killing vectors $(K^z_a, K^{\bar z}_a)$ and the CKV's $(C^z_a, C^{\bar z}_a)$, it is significant that, in the context of our non-compact AdS$_2$ geometry, they are very different geometrically. The former respect the boundary condition at $z{\bar z}=1$, while the latter do not. Indeed, the Killing vectors do not change anything at all: they are infinitesimal versions of the finite M\"{o}bius transformations that take $z\to \frac{az + b}{cz+d}$ (with ${\bar z}$ transforming analogously and $ad-bc=1$). 
In contrast, the CKV's that are not Killing Vectors change the metric by a non-trivial conformal factor; they are special only in that they preserve conformal gauge. Critical string theory is constructed such that the CKV's are elevated to symmetry generators, by other sectors of the theory compensating for the change in conformal factor. It would be interesting to contemplate the role of AdS$_2$ in the setting of critical string theory but we do not pursue this research direction in this article.    

In the JT-gravity approach to AdS$_2$ \cite{Almheiri:2014cka,Forste:2017apw,Forste:2017kwy,Engelsoy:2016xyb,Grumiller:2017qao,Cardenas:2018krd,Stanford:2019vob,Sachdev:2019bjn,Maldacena:2016upp,Saad:2019lba,Saad:2018bqo}, the geometry is regulated by a cut-off, and then the normalization conditions we impose in the strict AdS$_2$ limit do not apply.
This distinction does not matter for the CKV's; either way, they do not satisfy the applicable boundary conditions. However, in the regulated geometry, Killing vectors are legitimate symmetries so, in the context of a gravitational theory, configurations related by the $\SL(2,{\mathbb R})$ they generate are recognized as physically identical. This symmetry is gauged. In contrast, in the complete AdS$_2$, the configurations formally related by $\SL(2,{\mathbb R})$ are distinct, they are not identified, so the theory treats one of them as preferential. Thus, the symmetry is spontaneously broken.

In conclusion of this subsection: AdS$_2$ in conformal gauge does not support any normalizable Killing vectors and also no Conformal Killing vectors. 
In analogy with the standard world-sheet terminology we conclude that, after gauge fixing, there are {\it no residual gauge symmetries}. This result differs from the familiar analysis based on JT-gravity, partially summarized in the preceding paragraph. This is possible because we recognize the entire AdS$_2$ spacetime $|z|<1$, without imposing any cut-off, and the conformal factor in AdS$_2$ diverges as $|z|\to 1$.

\subsection{Conformal Gauge and Large Diffeomorphisms}

In presentations of world-sheet string theory, it is customary to exploit from the outset the freedom to reparametrize two coordinates, and so fix the world-sheet metric to conformal gauge \cite{Polchinski:1998rq}. It is always understood, but often not stressed at the outset, that for all Riemann surfaces, except $S^2$, this maneuver fails to remove a finite number of parameters known as {\it Quadratic Holomorphic Differentials} (QHD's). For compact Riemann surfaces, the ``would-be" diffeomorphisms that could formally enforce conformal gauge violate periodicity conditions, and the QHD's parametrize a space of configurations that are physically distinct because such periodicities differs.

In the non-compact AdS$_2$ geometry there are {\it infinitely} many parameters that cannot be removed by the freedom to reparametrize two coordinates. The 
``would-be" diffeomorphisms that could formally transform the geometry into conformal gauge are non-normalizable, so they cannot be used for gauge fixing. Alas, these ``would-be" symmetry generators parametrize a physical configuration space. 

We illustrate this point with a simple example.
Consider a class of metrics that are not in conformal gauge:
\begin{align}
	ds^2 = \frac{4dzd\zb}{(1-z\zb)^2} + c_n n(n^2-1) z^{n-2} dz^2 + \bar c_n n(n^2-1) \zb^{n-2} d\zb^2
	\ .\label{ads2_compl2}
\end{align}
Here $n\geq 2$ is an integer, $c_n$ is a complex number, and $\bar c_n$ is its complex conjugate. Here, and in the remainder of this subsection, we work only to linear order in $c_n$ and $\bar c_n$. Explicit computation shows that the curvature of the general metric \eqref{ads2_compl2}
is constant and negative, so the geometry is locally AdS$_2$
for any $c_n, {\bar c}_n$. 

Transforming the metric \eqref{ads2_compl2} into conformal gauge \eqref{ads2_compl} amounts to finding a diffeomorphism that removes the $c_n, {\bar c}_n$. 
Equivalently, we must solve the following system of differential equations for the generating vector field $\xi^{(n)}$,
\begin{align}
    2\nabla_z \xi^{(n)}_{z} = -c_n n(n^2-1) z^{n-2}
    \ ,\quad
    2\nabla_\zb \xi^{(n)}_{\zb} = -\bar c_n n(n^2-1) \zb^{n-2}
    \ ,\quad
    \nabla_{z} \xi^{(n)}_{\zb} + \nabla_{\zb} \xi^{(n)}_{z} = 0
    \ .
\end{align}
There is an exact solution with components \cite{Choi:2021nnq}
\begin{align}
	\xi^{(n)z}
	&=
		-\frac{c_n}{2}z^{n+1}
        + \frac{\bar c_n}{4}\Big(
    		n(n+1)\zb^{n-1}
    		- 2(n^2-1) z \zb^n
    		+ n(n-1) z^2\zb^{n+1}
    	\Big)
	\ ,\nonumber\\
	\xi^{(n)\zb}
	&=
		-\frac{\bar c_n}{2}\zb^{n+1}
        + \frac{c_n}{4}\Big(
    		n(n+1)z^{n-1}
    		- 2(n^2-1)\zb z^n
    		+  n(n-1) \zb^2 z^{n+1}
    	\Big)
    \label{xi2}
    \ .
\end{align}
If taken at face value, this result exhibits a diffeomorphism that takes the general metric \eqref{ads2_compl2} to conformal gauge \eqref{ads2_compl}. This would conform with the na\"{\i}ve standard world-sheet theory where it is sufficient to consider metrics in conformal gauge on the grounds that \textit{any} metric that is not already in conformal gauge is diffeomorphic to one that is. 

As noted already, this reasoning fails for Riemann surfaces with genus $g\geq 1$, and it also fails for the AdS$_2$ 
metric \eqref{ads2_compl2}. The ``gauge-fixing'' vector fields $\xi^{(n)}$ are formal;  they are not normalizable:
\begin{align}
    \int dzd\zb \sqrt g |\xi^{(n)}|^2\ =\ \infty
    \ ,
\end{align}
because the metric component $g_{z\zb}$ diverges as $|z|\to 1$.
We must therefore reject these candidate diffeomorphisms, just as we previously showed that, in AdS$_2$, there are no genuine Killing vectors and no CKV's.
Because of this, the two metrics \eqref{ads2_compl2} and \eqref{ads2_compl} are not gauge-equivalent; they parametrize physically distinct geometries.
By na\"{\i}vely working in conformal gauge, one misses out on a class of geometries that includes \eqref{ads2_compl2}.

\subsection{The Configuration Space of Distinct AdS\texorpdfstring{$_2$}{} Geometries}

The AdS$_2$ metrics \eqref{ads2_compl2} cannot be gauge-fixed to conformal gauge, so such metrics describe AdS$_2$ geometries that are physically inequivalent to the conformal disc \eqref{ads2_compl}. In these  basic examples the QHD's are simple powers $z^{n-2}dz^2$ and \eqref{xi2} are the vector fields that generate them \cite{Polchinski:1998rq,Larsen:2014bqa}.
In this subsection we aim to find the most general examples and characterize the space of such geometries

The basic example \eqref{ads2_compl2} pertains to a single value $n\geq 2$, but it can be consolidated by summing over all $n$. It follows that given an \textit{infinitesimal} holomorphic regular function $\ep(z)$ that satisfies $\ep(0)=\ep'(0)=\ep''(0)=0$, the following class of AdS$_2$ metrics are physically distinct:
\begin{align}
    ds^2 = \frac{4dzd\zb}{(1-z\zb)^2} + \ep'''(z) dz^2 + \bar\ep'''(\zb)d\zb^2
    \label{ads2_compl3}
    \ .
\end{align}
The antiholomorphic function $\bar\ep(\zb)$ is the complex conjugate of $\ep(z)$.
The diffeomorphism $\xi$ that maps \eqref{ads2_compl3} to the conformal disc \eqref{ads2_compl} is obtained by summing over \eqref{xi2} for $n\geq 2$ \cite{Choi:2021nnq},
\begin{align}
    \xi^z = -\frac{1}{2}\left(\ep(z)-\frac{1}{2}(1-z\zb)^2\bar\ep''(\zb)-z(1-z\zb)\bar\ep'(\zb)-z^2\bar\ep(\zb)\right)
    \ ,\nonumber\\
    \xi^\zb = -\frac{1}{2}\left(\bar\ep(\zb)-\frac{1}{2}(1-z\zb)^2\ep''(z)-\zb(1-z\zb)\ep'(z)-\zb^2\ep(z)\right)
    \label{xi3}
    \ .
\end{align}
This is the general large diffeomorphism at linear order.

One may readily see that this vector field is not normalizable in AdS$_2$.
Thus, each metric of the form \eqref{ads2_compl3} must be considered as parametrizing a physically distinct geometry from the conformal disc \eqref{ads2_compl}.
Moreover, two metrics of the form \eqref{ads2_compl3} with different functions $\ep_1(z)$ and $\ep_2(z)$ parametrize distinct geometries.
To see this, let $\xi_1$ and $\xi_2$ be the (non-normalizable) vector fields that generate large diffeomorphisms mapping each metric to the conformal disc.
Then the composite vector field $\xi_1-\xi_2$ maps one metric to the other, but it is not normalizable in AdS$_2$.

We have found that, at the infinitesimal level, there is a holomorphic \textit{ function}'s worth of AdS$_2$ geometries \eqref{ads2_compl3} that are distinct from one another as well as from the conformal disc \eqref{ads2_compl}.
We would like to do better and integrate this up to a \textit{finite} function's worth of metrics. The triple derivative of an infinitesimal function 
suggests a relation to the Schwarzian derivative.
Unfortunately, the straightforward replacement of $\ep'''(z)$ in \eqref{ads2_compl3} 
with a Schwarzian derivative gives a geometry that is no longer AdS$_2$.

To be more systematic, we consider the action of a general diffeomorphism $z\to Z(z,\zb)$ (and $\zb \to \Zb(z,\zb)$) on the conformal disc \eqref{ads2_compl}, and ask 
what conditions the complex function $Z(z,\zb)$ must satisfy for the resulting metric to be an integrated form of \eqref{ads2_compl3}. 
First, in the infinitesimal limit $Z(z,\zb) \approx z - \xi^z$ one should recover \eqref{xi3}.
Since the vector field $\xi$ in \eqref{xi3} is not normalizable, this makes $Z(z,\zb)$ a large diffeomorphism.
Second, the resulting metric should share the singularity structure of the conformal disc at $|z|= 1$, that is, it should take the form
\begin{align}
	ds^2 = \frac{4dzd\zb}{(1-z\zb)^2} + \text{(terms regular at $|z|\to 1$)}
	\label{complex_falloff}
	\ .
\end{align}
As an ansatz, we parametrize $Z(z,\zb)= z e^{C(z,\zb)}$, where $C=C(z,\zb)$ is some complex function of both $z$ and $\zb$.
By letting $zC$ to take the form of $\xi^z$ given in \eqref{xi3},
\begin{align}
	C(z,\zb) &=
		\frac{1}{2z}\left(
			\ep(z) - z^2\bar\ep(\zb) - z(1-z\zb)\pb\bar\ep(\zb) - \frac{1}{2}(1-z\zb)^2 \pb^2\bar\ep(\zb)
		\right)
	\ ,
    \label{C}
\end{align}
where $\ep(z)$ is no longer required to be infinitesimal, we achieve the first condition and obtain the correct infinitesimal form $Z(z,\zb)\approx z-\xi^z$.
The second condition \eqref{complex_falloff} puts fall-off conditions on the asymptotic behavior of $C$ near $|z|\to 1$:
\begin{gather}
	\pb C = \O((1-z\zb)^2)
	\ ,\qquad
	\Re C = \O(1-z\zb)
	\ ,\\
	e^{2\Re C}(1+z\p C)(1+\zb \pb \bar C)
	= \left(e^{2\Re C}+ \frac{1-e^{2\Re C}}{1-z\zb}\right)^2 + \O((1-z\zb)^2)
	\ .
\end{gather}
It turns out that \eqref{C} satisfies all these conditions.

Therefore, the \textit{finite} version of the metric \eqref{ads2_compl3} is given by performing $z\to Z(z,\zb)=z^{C(z,\zb)}$ on the conformal disc \eqref{ads2_compl} with $C(z,\zb)$ given by \eqref{C}:
\begin{align}
	ds^2
	&=
		\frac{
			e^{2\Re C}
		}{
			\left(e^{2\Re C}+\frac{1-e^{2\Re C}}{1-z\zb}\right)^2
		}
        \left[
            \frac{4(|1+z\p C|^2+ |\zb \pb C|^2)}{(1-z\zb)^2}dzd\zb
    	- \left(
                (1+z\p C)\p^3 \ep\, dz^2
                + \text{c.c.}
            \right)
        \right]
    \label{ads2_compl4}
	.
\end{align}
In other words, there is a \textit{finite} holomorphic function $\ep(z)$ that parametrizes the space of these metrics.
While the metric \eqref{ads2_compl4} has the desired asymptotic behavior \eqref{complex_falloff}, the explicit form is not particularly illuminating.
But in the next section, we shall see that its gauge theory formulation gives rise to a Schwarzian effective action.

\subsection{Relation to Fefferman-Graham Gauge and Diff(\texorpdfstring{$S^1$}{})}\label{sec:holomorphic_coord}

The main result of this section has been that deformations of the canonical metric on the conformal disc \eqref{ads2_compl} that take the holomorphic form $h_{zz} = \epsilon^{\prime\prime\prime}(z)dz^2$ \eqref{ads2_compl3}
at leading order cannot be gauge-fixed to conformal gauge.
In section \ref{sec:AdS_2} we studied the analogous question for the AdS$_2$ geometry in Fefferman-Graham coordinates and found that the metrics \eqref{ds2_SchTan} must be considered inequivalent. In both cases the obstruction is that the ``diffeomorphisms" that would formally do the deed are ``large", they are non-normalizable, so they are inadmissible. It is expected that these results should express identical underlying geometry, but in different coordinate systems, {\it i.e.} in different gauges. The goal of this subsection is to confirm this expectation and give a precise relation between the results.  

Recall that the linchpin of the large diffeomorphisms \eqref{2d_finite_real} in Fefferman-Graham gauge is the reparametrization $\tau \to f(\tau)$ of Euclidean time near the boundary of AdS$_2$. 
Infinitesimally, the transformation is simply $f(\tau) = \tau + \vep(\tau)$ 
or $\delta\tau=\vep(\tau)$ for an infinitesimal periodic function $\vep(\tau)$. 
From the map \eqref{ztaurho} to complex coordinates, one sees that this corresponds to an infinitesimal phase shift $\delta z = iz\vep(\tau)$, $\delta \zb = -i\zb\vep(\tau)$ near asymptotic infinity $|z|\to 1$.
Without loss of generality, we consider a single Fourier mode $\vep(\tau) = c_n e^{in\tau}$, where $n \geq 2$ and $c_n$ is an infinitesimal  complex parameter, such that $\delta z = ic_n ze^{in\tau}$ and $\delta\zb = -i c_n\zb e^{in\tau}$.
We exclude the $\SL(2)$ modes $n\neq 0,\pm 1$, and also the negative modes $n \leq -2$ which we return to later.
According to the map \eqref{ztaurho}, $z\sim e^{i\tau}$ and $\zb\sim e^{-i\tau}$ near the asymptotic boundary, which implies that in this region we can trade $e^{in\tau}$ for $z^n$ and $\zb$ for $z^{-1}$.
Thus, the phase transformation can be thought of as
\begin{align}
    \delta z = ic_nz^{n+1}
    \ ,\qquad
    \delta\zb = -ic_nz^{n-1}
    \ .
    \label{eqn:delzzbar}
\end{align}
For the moment the variations of $z$ and $\zb$ are not complex conjugates, they are related such that $\delta ( z \zb) = 0 $. This amounts to a Dirichlet condition imposed so the conformal boundary remains at the coordinate position $|z|=1$. Another perspective is that it is the continuation to $n\geq 2$ of the Killing vectors \eqref{eqn:KVs}, rather than the Conformal Killing Vectors \eqref{eqn:CvsK}.

The transformations \eqref{eqn:delzzbar} do not by themselves satisfy the boundary conditions, since the metric component $\delta g_{zz}$ that it generates has singularities near $|z|\to 1$. Following the strategy from 
section \ref{sec:AdS_2} that
lead to the Fefferman-Graham gauge transformation \eqref{2d_finite_real}, we can add corrections at subleading orders in $(1-z\zb)$ to find a metric with the correct asymptotic behavior as $|z|\to 1$.
As a result, we find the vector field
\begin{align}
	\xi^z = ic_nz^{n+1}
	\ ,\qquad
	\xi^\zb = -ic_nz^{n-1}
		\left(
			1
			+ (n-1)(1-z\zb)
			+ \frac{1}{2}(1-z\zb)^2 n(n-1)
		\right)
	\ ,
    \label{xi2.5}
\end{align}
which generates the metric deformation
\begin{align}
	\delta g_{zz} = -2ic_n n(n^2-1) z^{n-2}
	\ ,\qquad
	\delta g_{z\zb} = \delta g_{\zb\zb} = 0
	\ ,
	\label{ds2_QHD}
\end{align}
that is regular through the disc $0\leq |z|<1$.

We have restricted our attention to the positive modes $\vep(\tau) = e^{in\tau}$, $n\geq 2$, but an analogous analysis applies to the negative modes $\vep(\tau) = e^{-in\tau}$.
The positive and negative modes combine to the vector field \eqref{xi2} that generates the large diffeomorphisms: adding \eqref{xi2.5} and its complex conjugate yields \eqref{xi2}, up to an inconsequential redefinition of the coefficient $c_n$.
Thus, the diffeomorphism \eqref{xi2} can be derived by 
a procedure analogous to the one for Fefferman-Graham gauge in section \ref{sec:AdS_2}: perform a time reparametrization $\tau\to f(\tau)$, and keep the metric deformation normalizable.

The relation between the holomorphic coordinates in this section and
the Fefferman-Graham gauge in the previous one can be made more explicit as follows.
There is a diffeomorphism generated by a \textit{normalizable} vector field that maps the metric \eqref{ads2_compl2} to one in Fefferman-Graham gauge.
This implies that the two metrics are physically equivalent.
To see this, we first rewrite the metric \eqref{ads2_compl2} in the Fefferman-Graham coordinates by the map \eqref{ztaurho},
\begin{align}
	ds^2
	&=
		\left(
			1
			+ 2\frac{\left(\tanh\frac{\rho}{2}\right)^n}{(\sinh\rho)^2}
				\mathrm{Re}(c_ne^{in\tau})
		\right) d\rho^2
		+ 4i\frac{\left(\tanh\frac{\rho}{2}\right)^n}{\sinh\rho}\mathrm{Im}(c_ne^{in\tau})
			d\rho d\tau
		\nonumber\\&\quad
		+ \left(
			(\sinh\rho)^2
			-2\left(\tanh\frac{\rho}{2}\right)^n \mathrm{Re}(c_ne^{in\tau})
		\right) d\tau^2
	\ .
\end{align}
Next, we consider a vector field $\zeta^{(n)}$ with the following (lower) components,
\begin{align}
	\zeta^{(n)}_\rho
	&=
		\frac{\mathrm{Re}(c_n e^{in\tau})}{n^2-1}\left(
		1
		- \frac{(n+\cosh\rho)}{\sinh\rho}\tanh^n\frac{\rho}{2}
		\right)
	\ ,\\
	\zeta^{(n)}_\tau
	&=
		-i\mathrm{Im}(c_n e^{in\tau})
		\Bigg[
			\frac{1}{n}(\sinh\rho)^2
			- \frac{n\cosh\rho\sinh\rho}{n^2-1}
			\nonumber\\&\qquad\qquad\qquad
			+ \left(\frac{2n^2-1+2n\cosh\rho+\cosh(2\rho)}{2n(n^2-1)}\right) \tanh^n\frac{\rho}{2}
		\Bigg]
	\ .
\end{align}
Its norm has the asymptotic behavior $|\zeta^{(n)}|^2\sim \O(e^{-4\rho})$, so it is normalizable in AdS$_2$.
The diffeomorphism generated by $\zeta^{(n)}$ maps the metric \eqref{ads2_compl2} to
\begin{align}
	ds^2
	&=
		d\rho^2
		+ (\sinh\rho)^2 d\tau^2
		- 2\mathrm{Re}(c_ne^{in\tau}) e^{-\rho}\sinh\rho d\tau^2
	\ .
\end{align}
This is precisely the Fefferman-Graham metric \eqref{ds2_SchTan_infinitesimal} for an infinitesimal Euclidean time reparametrization $\delta \tau = \vep(\tau)$, where the parameter $\vep(\tau)$ is identified as
\begin{align}
	\vep(\tau) = -\frac{1}{2n(n^2-1)}\left((c_n+\bar c_n)\sin(n\tau)-i(c_n-\bar c_n)\cos(n\tau)\right)
	\ ,
\end{align}
in terms of the complex parameter $c_n$.

Summing up the powers $n\geq 2$, the analogous result applies to the metric \eqref{ads2_compl3} in terms of the holomorphic function $\ep(z)$. Thus, 
there is a {\it normalizable} vector field $\zeta$, the linear combination of $\zeta^{(n)}$ for all $n\geq 2$, that maps the conformal disc \eqref{ads2_compl3} to the Fefferman-Graham gauge \eqref{ds2_SchTan_infinitesimal}, with the infinitesimal parameter $\vep(\tau)$ identified in terms of $\ep(z)$ as
\begin{align}
	\vep(\tau) = \mathrm{Re}\left(ie^{-i\tau}\ep(e^{i\tau})\right)
	=\frac{i}{2}\left(e^{-i\tau}\ep(e^{i\tau}) - e^{i\tau}\bar\ep(e^{-i\tau})\right)
	\ .
    \label{vep_ep}
\end{align}
The vector field $\zeta$ is a composition $\zeta = v + \xi$ of two vector fields:
the large diffeomorphism $\xi$ \eqref{xi3} that generates holomorphic deformations, and its Fefferman-Graham analogue $v=\delta\tau \p_\tau + \delta\rho \p_\rho$ with components \eqref{diff_v} and $\vep(\tau)$ identified as \eqref{vep_ep}.
While both of $v$ and $\xi$ are non-normalizable in AdS$_2$, their composition $\zeta$ is normalizable and a conventional physical symmetry. 

In conclusion, the physically distinct geometries \eqref{ds2_SchTan_infinitesimal} in the Fefferman-Graham gauge are precisely the geometries \eqref{ads2_compl3} that are \textit{not} captured by the conformal gauge.

\section{The Gauge Theory Formulation of AdS\texorpdfstring{$_2$}{} Gravity}
\label{sec:gaugethy}

It is well-known that two-dimensional gravity can be cast as a topological gauge theory in its first-order formulation \cite{Isler:1989hq,Chamseddine:1989wn}.
In this section we use the BF theory \cite{Witten:1991we,Witten:1992xu} description of JT gravity 
to parametrize the space of physically distinct AdS$_2$ geometries. 

Recent discussions of the
BF approach to JT gravity focus on fields near the asymptotic boundary of AdS$_2$ \cite{Grumiller:2017qao,Freidel:2020xyx,Freidel:2020svx,Saad:2019lba,Heydeman:2020hhw}. We consider the entire spacetime, without restriction to
the boundary region.
We show that the effective action of the theory, given by a 1D $\sigma$-model, reduces to the Schwarzian action.

\subsection{BF Theory and JT Gravity}

The field content of BF theory consists of a gauge one-form $A$ and a scalar field $B$, both in the adjoint representation of the gauge group.
The action of the theory takes the gauge-invariant form
\begin{align}
	S_\text{JT} = -i \int \Tr(B F)
	\ ,
	\label{BFJT}
\end{align}
where $F=dA-A\wedge A$ is the field strength.
The equations of motion for the fields $A$ and $B$ are obtained by varying the action \eqref{BFJT} with respect to $B$ and $A$ respectively:
\begin{align}
	F = dA -A\wedge A = 0
    \label{bf_eom_A}
	\ ,\\
	DB=dB-[A,B]=0
    \label{bf_eom_B}
	\ .
\end{align}
The gauge covariant derivative $D$ is 
defined by \eqref{bf_eom_B}.

The fields decompose into a Lie algebra basis as $A(x) = A^i(x) P_i$ and $B(x) = B^i(x) P_i$. 
We are interested in the $\sl(2,\R)$ algebra where
the basis elements satisfy 
\begin{align}
	[P_0, P_1] = P_2
	\ ,\qquad
	[P_0, P_2] = -P_1
	\ ,\qquad
	[P_1, P_2] = -P_0
	\ .
	\label{P}
\end{align}
We employ the following explicit matrix representations for $P_i$:
\begin{align}
	P_0 = \frac{1}{2}\begin{pmatrix}
		0&1\\-1&0
	\end{pmatrix}
	\ ,\qquad
	P_1 = \frac{1}{2}\begin{pmatrix}
		-1&0\\0&1
	\end{pmatrix}
	\ ,\qquad
	P_2 = \frac{1}{2}\begin{pmatrix}
		0&1\\1&0
	\end{pmatrix}
	\ .
\end{align}
Thus the fields $A$ and $B$ are written as traceless real matrices.
The Cartan metric reads $2\Tr(P_i P_j) = \eta_{ij} = \diag(-1,1,1)$. Note that
in our notation $i,j,\ldots\in\{0,1,2\}$ 
comprise all three $\SL(2,\R)$ indices. 

BF theory with $\SL(2,\R)$ gauge group is a first-order formulation of JT 
gravity \cite{Mertens:2018fds,Iliesiu:2019xuh,Heydeman:2020hhw}.
The identification encodes first and second order aspects of 2D gravity in the gauge field $A$. Its ``spatial" components correspond to the 2D zweibein via $A^a(x) = e^a(x)$ where $a,b,\ldots\in\{1,2\}$.
The ``temporal" component of the gauge field is identified with the spin connection\footnote{Here $\w$ is the Hodge dual of the spin connection $\w^a{}_b = \ep^a{}_b \w$, with the Levi-Civita tensor $\ep^{12}=\ep_{12}=\ep^1{}_2=1$.} 
through $A^0(x) = \w(x)$. With these identifications, expansion of the field strength in our basis for the Lie algebra gives
\begin{align}
	F
	&=
		(d\w + e^1\wedge e^2) P_0
		+ (de^a + \w^a{}_b\wedge e^b)P_a
	\ .
	\label{FBF}
\end{align}
Inserting this expression into the bulk action \eqref{BFJT}, we find its component form: 
\begin{align}
	S_\text{BF}
	&=
		-\frac{i}{2}\int \left[
			B^0 (d\w + e^1\wedge e^2)
			+ B^a (de_a + \w_{ab} \wedge e^b)
		\right]
	\ .
	\label{BFJT2}
\end{align}

The BF action in its original form 
\eqref{BFJT} is manifestly invariant under
$\SL(2,\R)$ gauge symmetry. The  
infinitesimal gauge transformation
$\delta A = D\ep = d\ep - [A,\ep]$ is parametrized by an element $\epsilon(x) = \epsilon^i(x)P_i$ of $\sl(2,\R)$. Among these, the ``spatial" 
parameters $\ep^a$ generate the diffeomorphism $\xi^\mu = \ep^a e_a{}^\mu$,
up to the equations of motion. 
Thus the component form of the BF action \eqref{BFJT2} is a gravitational action; it is invariant under 2D diffeomorphism symmetry. 

The third component of the $\sl(2,\R)$ symmetry
is proportional to the parameter $\ep^0$ and  generates rotations of the Lorentz frame. 
This additional invariance under local rotation of Lorentz frame is a well-known feature of the first order formalism, sometimes known as the Palatini formalism. 
It appears here because, in the original gauge theory formulation 
\eqref{BFJT}, the field components
\begin{align}
    A^0 = \w~,\qquad A^a = e^a
    \label{eqn:Awe}
    \ ,
\end{align}
are independent variables. Therefore, after expansion in components and a change in notation, the zweibein $e^a$ and spin connections $\w$ remain independent variables.

The two scalar fields $B^a$ act as Lagrange multipliers to impose the vanishing torsion constraints $T^a=de^a+\w^a{}_b\wedge e^b=0$.
Thus the spin connections $\w^a{}_b$ 
which, on a 2D manifold, reduce to the single component $\w$, are determined from the zweibein via the usual Cartan equations. 
We impose this constraint directly in the action \eqref{BFJT2} and note that in two dimensions we have $d\w = \frac{1}{2} {\cal R} e^1\wedge e^2$ where ${\cal R}$ is the Ricci scalar. This gives the bulk action
\begin{align}
	S_\text{BF}
	&=
		-\frac{i}{4}\int d^2x \sqrt g\,
			B^0 ({\cal R} + 2)
	\ ,
 \label{eqn:JT2ndorder}
\end{align}
where we used $e^1 \wedge e^2 = \sqrt g d^2x$.
This is the JT gravity action \cite{Jackiw:1982hg,Jackiw:1984je,Teitelboim:1983ux}, with the scalar $B^0(x)$ playing the role of the JT-dilaton that some references denote $\Phi(x)$. The equation of motion of this scalar field ensures that the geometry has constant negative curvature 
${\cal R}=-2$, {\it i.e.}
it is AdS$_2$, at least locally.

\subsection{Large Gauge Transformations}
In sections \ref{sec:AdS_2} and \ref{sec:worldsheet}, we interpreted the space of ``all" AdS$_2$ geometries in the sense of general relativity, {\it i.e.} as the space of distinct metrics modulo normalizable (!) diffeomorphisms. BF theory is more abstract, in the sense that there is no geometry {\it a priori}. However, it is based on a particularly simple principle: we must consider all distinct gauge field configurations, modulo normalizable (!) gauge transformations. It is instructive to show that {\it all} these principles are consistent with one another, in that they give the same configuration space. 

In fact, at first sight, it may appear that the second order form of the BF action
\eqref{eqn:JT2ndorder}, {\it i.e.} the JT gravity action in its standard form, could not possibly describe AdS$_2$ quantum gravity by itself, because 
the JT scalar $B^0(x)$ appears explicitly in the action \eqref{eqn:JT2ndorder}.
However, as we have emphasized throughout, we study AdS$_2$ quantum gravity in its non-compact form, without a regulating cut-off. In this setting there are no normalizable scalar modes, so $B^0(x)$ is not a dynamical
field. Comparing its equation of motion 
$$
(\nabla_\mu \nabla_\nu - g_{\mu\nu})B^0(x)=0~,
$$
with \eqref{eqn:Xeqn}, we can identify $B^0$ with one of the potentials $X_{\pm 1}, X_0$, introduced in subsection \ref{subsec:confgauge} where they appeared because 
their curl and gradients yield Killing vectors and CKV's, respectively. The (C)KV's obtained this way are not normalizable and the scalar fields that generate them are even worse: the fields themselves diverge as $|z|\to 1$, and the AdS$_2$ volume only aggravates the problem. Therefore, in our approach, excitations of the scalar field is 
$B^0$ is not normalizable. Since we do not regularize AdS$_2$, it is not even possible that a ``renormalized" scalar remains.



Working with a gauge field instead of a metric has several advantages. For example, some technical manipulations simplify: 
diffeomorphisms become matrix multiplications, and thus finding inverse transformations is nearly trivial.
Contrast this to the large diffeomorphisms \eqref{2d_finite_real} in Fefferman-Graham gauge, whose inverse transformation is much more complicated.
Also, in gauge theory the connection between field configurations and the effective action of spontaneously broken symmetry becomes more transparent: it is given by the $\sigma$-model action for particle
motion on a simple (but non-compact) group.

To study the space of distinct gauge field configurations and characterize large gauge transformations, we solve the equations of motion \eqref{bf_eom_A} for $A$.
Since the theory is a first-order formulation of JT gravity \eqref{eqn:JT2ndorder}, the space of solutions are analogues of the AdS$_2$ geometries \eqref{ds2_SchTan}. The gauge field configurations obtained via the identification \eqref{eqn:Awe} are: 
\begin{align}
	A
	&=
		\frac{1}{2}\begin{pmatrix}
			-1 &0\\0&1
		\end{pmatrix}
		d\rho
		+ \frac{1}{2} \begin{pmatrix}
			0& -2e^{-\rho}\Sch\left(\tan\frac{f(\tau)}{2},\tau\right)\\e^\rho&0
		\end{pmatrix}
		d\tau
		\label{BF_solution_A}
	\ .
\end{align}
There is one gauge field configuration for each function $f(\tau)$ on the circle satisfying the constraints \eqref{eqn:f_conditions}.
For a given function $f(\tau)$, and thus a gauge field $A$ \eqref{BF_solution_A}, we solve the equation of motion \eqref{bf_eom_B} for the adjoint scalar field $B$ and find
\begin{align}
	B &=
	\begin{pmatrix}
		\beta'(\tau) & -2e^{-\rho}\left[\beta(\tau)\Sch\left(\tan\frac{f(\tau)}{2},\tau\right)+\beta''(\tau)\right] \\
		e^\rho \beta(\tau) & -\beta'(\tau)
	\end{pmatrix}
		\label{BF_solution_B}
	\ ,
\end{align}
where $\beta(\tau)$ is the function 
\begin{align}
	\beta(\tau)
	\ &=\ 
		\frac{1}{\p_\tau \tan\frac{f(\tau)}{2}}\left(
			\beta_0
			+ \beta_1\tan\frac{f(\tau)}{2}
			+ \beta_2\tan^2\frac{f(\tau)}{2}
		\right)
	\ ,
\label{BF_solution_b}
\end{align}
defined by $\tan\frac{f(\tau)}{2}$ and three integration constants $\beta_0$, $\beta_1$, and $\beta_2$.
The expressions (\ref{BF_solution_A}-\ref{BF_solution_b}) are exact in that they are valid to all orders in $e^{-\rho}$.

The JT gravity interpretation of BF theory maps each gauge field configuration \eqref{BF_solution_A} to an AdS$_2$ geometry via the map \eqref{eqn:Awe}.
Therefore, gauge transformations that leave the gauge field \eqref{BF_solution_A} invariant are ``isometries'' of the corresponding metric 
Such gauge transformations are, infinitesimally, parametrized by $\sl(2,\R)$ matrices $\lambda$ that satisfy $\delta_\lambda A = D\lambda = 0$.
This is precisely the equation of motion \eqref{bf_eom_B}, with $B$ being the matrix $\lambda$.
Given a geometry with a fixed function $f(\tau)$, its space of solutions \eqref{BF_solution_B} is a three-dimensional vector space generated by the three parameters $\beta_0$, $\beta_1$ and $\beta_2$.
This space corresponds to the space of scalar potentials generated by $X_a$, introduced  in subsection \ref{subsec:confgauge}, whose curls yield the formal AdS$_2$ Killing vectors.

As solutions to $F=0$ the configurations \eqref{BF_solution_A} are flat connections and 
so they are formally pure gauge. In other words, 
for each function $f(\tau)$
there exists a group element $g\in\SL(2,\R)$  such that $A = (dg) g^{-1}$.
As a reference, we introduce the gauge connection 
$A_0$ that corresponds to the identity function $f(\tau)=\tau$. For this reference configuration the Schwarzian for $f(\tau)=\tau$ is a constant $\Sch(\tau)=\frac{1}{2}$ and so
\begin{align}
	A_0
	&=
		\frac{1}{2}\begin{pmatrix}
			-1 &0\\0& 1
		\end{pmatrix}
		d\rho
		+ \frac{1}{2} \begin{pmatrix}
			0& -e^{-\rho}\\ e^\rho&0
		\end{pmatrix}
		d\tau
	\ .
 \label{eqn:A0conf}
\end{align}
This is formally pure gauge $A_0=(dg_0)g_0^{-1}$ where the group element $g_0$ is
\begin{align}
	g_0 = \begin{pmatrix}
		e^{-\rho/2}\cos\frac{\tau}{2} & -e^{-\rho/2}\sin\frac{\tau}{2} \\
            e^{\rho/2}\sin\frac{\tau}{2} & e^{\rho/2}\cos\frac{\tau}{2}
	\end{pmatrix}
	\ .
    \label{eqn:g0}
\end{align}

Starting from the reference configuration $A_0$, 
the configuration \eqref{BF_solution_A} for a general function $f(\tau)$ is achieved by a gauge transformation $g_f$ that 
solves 
$$
g_fA_0g_f^{-1} + (dg_f)g_f^{-1} = A~.
$$
It has the explicit form
\begin{align}
	g_f=\begin{pmatrix}
		a(\tau) & e^{-\rho}b(\tau)\\e^\rho c(\tau) & d(\tau)
	\end{pmatrix}
	\ ,
	\label{eqn:gf}
\end{align}
where the functions $a$, $b$, $c$ and $d$ are given by
\begin{align}
	a(\tau)
	&=
		\frac{1}{f'(\tau)^{3/2}}\left[
			f'(\tau)^2\cos\left(\frac{f(\tau)-\tau}{2}\right)
			- f''(\tau)\sin\left(\frac{f(\tau)-\tau}{2}\right)
		\right]
	\ ,\\
	b(\tau)
	&=
		-\frac{1}{f'(\tau)^{3/2}}\left[
			f'(\tau)^2\sin\left(\frac{f(\tau)-\tau}{2}\right)
			+ f''(\tau)\cos\left(\frac{f(\tau)-\tau}{2}\right)
		\right]
	\ ,\\
	c(\tau)
	&=
	\frac{1}{\sqrt{f'(\tau)}}\sin\left(\frac{f(\tau)-\tau}{2}\right)
	\ ,\\
	d(\tau)
	&=
	\frac{1}{\sqrt{f'(\tau)}}\cos\left(\frac{f(\tau)-\tau}{2}\right)
	\ .
\end{align}
Thus the general flat connections \eqref{BF_solution_A} are related to the reference configuration \eqref{eqn:A0conf} via the $\SL(2,\mathbb{R})$ gauge transformations \eqref{eqn:gf}, at least formally. 

As we have emphasized repeatedly in previous sections, when {\it local} symmetries, such as gauge transformations,  are normalizable, they {\it identify} configurations that appear distinct. On the other hand, non-normalizable ``gauge-transformations" map physically distinct configurations to one another.  

In the case at hand, an infinitesimal gauge transformation $\Lambda\in\sl(2,\R)$ is normalizable when the
$\SL(2,\R)$ invariant norm:
\begin{align}
	\int \Tr(B d\Lambda\wedge d\Lambda) < \infty
	\label{BF_norm_Lambda}
	\ .
\end{align}
The analogous condition for a finite gauge transformation $g=e^\Lambda$ is
\begin{align}
	\int  \Tr(B d g\wedge dg^{-1}) < \infty
	\label{BF_norm_g}
	\ .
\end{align}
According to this criterion, the gauge element $g_f$ \eqref{eqn:gf} is {\it not} normalizable because the expression
\begin{align}
	\Tr\left(B (\p_\rho g_f) g_f^{-1}(\p_\tau g_f) g_f^{-1}\right)
	\ ,
\end{align}
with $B$ in \eqref{BF_solution_B}, turns out to be independent of $\rho$, so the
integral over $d\rho\wedge d\tau$ diverges.
Therefore, the group transformations \eqref{eqn:gf}
map the reference point $A_0$ to configurations \eqref{BF_solution_A} that must be considered physically distinct. 

The conclusion is that, in the BF-formalism, the space of physically distinct configurations are parametrized by a single function $f(\tau)$.
This result agrees with the more conventional gravity analysis, where the space of metrics 
in Fefferman-Graham gauge \eqref{ds2_SchTan} are parametrized by a function $f(\tau)$ on the circle satisfying \eqref{eqn:f_conditions}.

\subsection{Symplectic Structure}

In this subsection, we examine the symplectic structure on the configuration space \cite{Witten:1991we,Witten:1992xu}.

As discussed in the previous subsection, the configuration space is a space of flat $\SL(2,\R)$ gauge connections over Euclidean AdS$_2$, up to redundant gauge transformations.
The variation $\delta A$ at each point $A$ is a tangent vector on this configuration space (and a one-form in spacetime) that transforms in the adjoint representation of $\SL(2,\R)$.
Thus, given two variations $\delta_1 A$ and $\delta_2 A$, the space has a natural symplectic structure defined by the symplectic form $\Omega$ \cite{Witten:1991we,Saad:2019lba}
\begin{align}
	\Omega(\delta_1 A,\delta_2 A) = \gamma\int \Tr (\delta_1 A \wedge \delta_2 A)
	\ ,
	\label{Omega}
\end{align}
where $\gamma$ is some dimensionful constant.
The domain of the integral in \eqref{Omega} is the 2D spacetime.
The symplectic form $\Omega$, being a two-form in configuration space, maps two tangent vectors $\delta_1A$ and $\delta_2A$ to a number, the r.h.s$.$ of \eqref{Omega}.

In our context the tangent vectors $\delta_iA$ ($i=1,2$) are formally pure gauge, so $\delta_i A = d\Lambda_i - [A,\Lambda_i]$ for some $\sl(2,\R)$ matrices $\Lambda_i$ and then
\begin{align}
    \Tr\Big(
        (d\Lambda_1-[A,\Lambda_1]) \wedge (d\Lambda_2-[A,\Lambda_2])
    \Big)
    &=
        d\Tr(\Lambda_1(d\Lambda_2-[A,\Lambda_2]))
        - \Tr(F[\Lambda_1,\Lambda_2])~.
\end{align}
This identity is for two pure gauge deformations of any gauge field configuration. For flat backgrounds $F=0$
the expression is $d$ of something and then the 2D integral in the symplectic form \eqref{Omega} reduces to a 
1D boundary integral
\begin{align}
    \Omega = \gamma\int\Tr(\Lambda_1(d\Lambda_2-[A,\Lambda_2]))
    \ .
    \label{eqn:Omega1D}
\end{align}

The tangent vectors $\delta_1 A$ and $\delta_2A$ relevant for our configuration space are non-normalizable variations of the gauge field
	that keep the form of \eqref{BF_solution_A} fixed,
	but changes the Schwarzian by a function composition $f \to f\circ g$.
The change in function $f$ under a composition reads $\delta f = \delta(f\circ g)|_{g=\id}=f'(\tau)\delta g$.
Using this, we obtain from \eqref{BF_solution_A} an explicit expression for the tangent vector $\delta A$,
\begin{align}
	\delta A
	&=
		\begin{pmatrix}
			0 & -e^{-\rho}
                \left[
		\left(
            2\delta g'(\tau)
		  + \delta g(\tau) \p_\tau
        \right)\Sch\left(\tan\frac{f(\tau)}{2},\tau\right)
		+ \delta g'''(\tau)
                \right]d\tau
                \\
                0 & 0
		\end{pmatrix}
	\ .
    \label{eqn:deltaA}
\end{align}
The terms in square brackets follow from the composition law of Schwarzian derivatives.
This expression is a formally pure gauge $\delta A = d\Lambda - [A,\Lambda]$ with the $\sl(2,\R)$ gauge function
\begin{align}
	\Lambda = 
	\begin{pmatrix}
		\frac{1}{2}\delta g'(\tau) & -e^{-\rho}\left[\delta g(\tau)\Sch\left(\tan\frac{f(\tau)}{2},\tau\right)+\delta g''(\tau)\right] \\
		\frac{1}{2}e^\rho\delta g(\tau) & -\frac{1}{2}\delta g'(\tau)
	\end{pmatrix}
	\ .
    \label{eqn:lambda}
\end{align}
Inserting into \eqref{eqn:Omega1D} two such gauge matrices 
$\Lambda_1$ and $\Lambda_2$ parametrized by $\delta g_1(\tau)$ and $\delta g_2(\tau)$ respectively, we obtain the expression
\begin{align}
	\Omega
	&=
		\frac{\gamma}{4}
		\int_0^{2\pi} d\tau \left[
			\delta g'(\tau)\wedge \delta g''(\tau)
			- 2\Sch\left(\tan\frac{f(\tau)}{2},\tau\right)\delta g(\tau)\wedge \delta g'(\tau)
		\right]
	\ ,
	\label{eqn:symp}
\end{align}
where $\wedge$ in \eqref{eqn:symp} represents wedge product in phase space: $\delta g\wedge \delta g' \equiv \delta g_1 \delta g_2' - \delta g_1' \delta g_2$ and $\delta g'\wedge \delta g'' \equiv \delta g_1' \delta g_2'' - \delta g_1'' \delta g_2'$.

Throughout this paper we emphasize a ``bulk" perspective, we discuss the configuration space of AdS$_2$ geometry. 
The mathematical fact that allow this to be identified with a ``boundary theory" is that it has
the same symplectic structure as the space of diffeomorphisms on a circle, modulo $\SL(2,\R)$ transformations \cite{Oblak:2016eij,Stanford:2017thb}.
The space $\Diff(S^1)/\SL(2,\R)$ can be represented as
the unique coadjoint orbit of the Virasoro group that has $\SL(2,\R)$ stabilizer \cite{Kirillov1981,Witten:1987ty,Bakas:1988bq}.
Any coadjoint orbit is a symplectic manifold, endowed with a natural symplectic form by the Kirillov-Kostant-Souriau (KKS) theorem \cite{Kirillov1974elements,Kostant1970orbits,Souriau1970structure}.
The corresponding symplectic form of the $\mathrm{Diff}(S^1)/\SL(2,\R)$ manifold takes the form (see \cite{Stanford:2017thb} for example),
\begin{align}
	\Omega_\text{KKS}
	&=
		\int_0^{2\pi} d\tau \left[
			\left(\frac{\rmd f}{f'}\right)'
			\wedge
			\left(\frac{\rmd f}{f'}\right)''
			- 2\Sch\left(\tan\frac{f(\tau)}{2},\tau\right)
				\left(
					\frac{\rmd f}{f'}
				\right)
				\wedge
				\left(
					\frac{\rmd f}{f'}
				\right)'
		\right]
	\ .
	\label{KKS}
\end{align}
The phase space of the theory is the abstract space of functions $f(\tau)$ on the unit circle.
The lowest three Fourier modes $f(\tau)= e^{in\tau}$ $(n=0,\pm 1)$ are zero modes that comprise the $\SL(2,\R)$ subgroup.
The phase space one-form $\rmd f$ is an element of the cotangent space of a point $f$ in the function space,
	and therefore corresponds to an infinitesimal variation of $f$.
Such variations can naturally be written in terms of function compositions $f\to f\circ g$ as $\rmd f = \rmd(f\circ g)|_{g=\id}= f'(\tau) \rmd g$.
Different functions $g$'s represent different tangent vectors at the point $f$ in phase space.
Inserting this into the KKS symplectic form \eqref{KKS}, we obtain
\begin{align}
	\Omega_\text{KKS}
	&=
		\int_0^{2\pi} d\tau \left[
			\rmd g'\wedge\rmd g''
			- 2\Sch\left(\tan\frac{f(\tau)}{2},\tau\right)
				\rmd g \wedge \rmd g'
		\right]
	\ .
\end{align}
Up to an overall factor, this is the symplectic form \eqref{eqn:symp} on the space of physically distinct gauge field configurations.

\subsection{The Effective Action}

The gauge field configurations \eqref{BF_solution_A} are mapped to one another by formal gauge transformations that are not normalizable.
Thus, they form a space of physically distinct solutions.
We now show that the effective field theory
describing this physical configuration space is governed by a Schwarzian action.

The gauge field \eqref{BF_solution_A} for $f(\tau)$ can be written in terms of a single group element $A= ( dg )g^{-1}$, where $g=g_fg_0$. The reference $g_0$ is fixed and arbitrary, in the sense that any other reference configuration would by physically equivalent. The $g_f$ represent departures away from the reference, via a motion through the configuration space of all matrices \eqref{eqn:gf} with functions $f(\tau)$ satisfying the conditions \eqref{eqn:f_conditions}.
Because this set-up is purely geometrical, the low energy effective theory is 
completely determined. It is governed by a 1D $\sigma$-model action proportional to
\begin{align}
    I_\text{eff}
    &=
        \int_0^{2\pi} d\tau \Tr\left[(g^{-1} \p_\tau g)^2\right]
    .
    \label{eqn:1d_sigma_action}
\end{align}
Using the explicit expressions \eqref{eqn:g0} and \eqref{eqn:gf}, it is straightforward to compute $g=g_fg_0$ and insert it into \eqref{eqn:1d_sigma_action} to find 
\begin{align}
    I_\text{eff}
    &=
        -\int_0^{2\pi} d\tau \Sch\left(\tan\frac{f(\tau)}{2},\tau\right)
	\ . 
\end{align}
These expressions should be furnished with an overall constant of proportionality 
that is dimensionful and sets the scale of the effective theory. This scale is arbitrary from the low energy effective field theory point of view but ultimately determined by matching with a theory that is valid at higher energy. 

The derivation of the Schwarzian action is particularly satisfying from the gauge theory point of view. It is unsurprising that a similar approach applies in the Fefferman-Graham formulation from section \ref{sec:AdS_2}, since the Schwarzian derivative appears explicitly in the geometries \eqref{ds2_SchTan} that represent the configuration space.

The situation in the holomorphic coordinates in section \ref{sec:worldsheet} is more complicated. We have demonstrated in detail that the subset of the geometries \eqref{ads2_compl3} that correspond to infinitesimal deformations of the conformal disc \eqref{ads2_compl} are gauge equivalent to the geometries \eqref{ds2_SchTan_infinitesimal} in Fefferman-Graham gauge. However, the Schwarzian derivative does not appear manifestly in the holomorphic coordinates so the 
effective theory of the full space of geometries \eqref{ads2_compl4} that we are unable to gauge-fix to a manifestly holomorphic gauge is less clear. 
In the following we close this gap by showing that, in gauge theory, these geometries yield a space of gauge configuration whose effective action is governed by the Schwarzian derivative of some function $f(\tau)$, as expected. 

According to the dictionary \eqref{eqn:Awe}, the conformal disc metric \eqref{ads2_compl} corresponds to the pure-gauge configuration $A_0=(dg_0)g_0^{-1}$ with the element
\begin{align}
	g_0 = \frac{1}{2(z\zb)^{1/4}}\begin{pmatrix}
		(\sqrt z+\sqrt \zb)\frac{\sqrt{1-z\zb}}{1+\sqrt{z\zb}}
		&
		i(\sqrt z-\sqrt \zb)\frac{\sqrt{1-z\zb}}{1+\sqrt{z\zb}}
		\\
		-i(\sqrt z-\sqrt \zb)\frac{1+\sqrt{z\zb}}{\sqrt{1-z\zb}}
		&
		(\sqrt z+\sqrt \zb)\frac{1+\sqrt{z\zb}}{\sqrt{1-z\zb}}
	\end{pmatrix}
	\ .
\end{align}
The diffeomorphism $z\to Z(z,\zb)=ze^{C(z,\zb)}$ with $C$ given in \eqref{C} is generated by a non-normalizable vector field.
It corresponds to the large gauge transformation $g_0 \to g_\ep$, which maps the point $A_0$ in configuration space to another point $A_\ep=(dg_\vep)g_\ep^{-1}$.
Near $|z|\to 1$, this becomes a phase transformation $\tau\to f(\tau)=\tau-\vep(\tau)$ of the holomorphic coordinate \eqref{ztaurho}, where $\vep(\tau)$ is given in terms of $\ep(z)$ and $\bar\ep(\zb)$ in \eqref{vep_ep},
\begin{align}
	f(\tau) = \tau-\Re\left[i e^{-i\tau}\ep(e^{i\tau})\right]
	\ .
\end{align}
By direct computation, one may see that
\begin{align}
           \Tr\left[(g_\ep^{-1}\p_\tau g_\ep)^2\right]
	=
		\frac{2Z\Zb(\p_\tau \Re C)^2}{(1-Z\Zb)^2}
		- \frac{1}{2} \left(1-\frac{i}{2}\p_\tau(C-\bar C)\right)^2
	\ ,
\end{align}
which is finite as $|z|\to 1$, since $\Re C\sim (1-z\zb)$, and
\begin{align}
	\frac{\p_\tau\Re C}{1-z\zb}
	&=
		\frac{i}{4z\zb}\left(
			- 2\zb\ep
			+ 2z\zb\p\ep
			- z(1+z\zb)\p^2\ep
			- z^2(1-z\zb)\p^3\ep
		\right)
		+ \text{c.c.}
	\ .
\end{align}
Evaluating at $|z|\to 1$, this simplifies to the following expression in terms of the function $f(\tau)$,
\begin{align}
	\left.\Tr\left[(g_\ep^{-1}\p_\tau g_\ep)^2\right]\right|_{|z|\to 1}
	&=
		\frac{f''(\tau)^2}{2f'(\tau)^2}
		-\frac{1}{2}f'(\tau)^2
	\ .
\end{align}
When integrating this over a circle, one can add in a total derivative to obtain
\begin{align}
	\int_0^{2\pi} d\tau \Tr\left[(g_\ep^{-1}\p_\tau g_\ep)^2\right]|_{|z|\to 1}
	&=
		-\int_0^{2\pi} d\tau \Sch\left(\tan\frac{f(\tau)}{2},\tau\right)
	\ .
\end{align}
Therefore, the effective action of the theory is governed by the Schwarzian derivative of the function $f(\tau)$.

\section{AdS\texorpdfstring{$_2$}{} from Dimensional Reduction of AdS\texorpdfstring{$_3$}{}}
\label{sec:AdS3toAdS2}
The holographic understanding of AdS$_{d+1}$/CFT$_d$ correspondence with $d=1$ has been recognized as a special and confusing case ever since the advent of the AdS/CFT correspondence \cite{Maldacena:1997re}. Conversely, AdS$_3$ and its dual CFT$_2$ is the simplest case in many aspects. In this section we discuss 
AdS$_2$ from an AdS$_3$ point of view.

\subsection{Large Diffeomorphisms in AdS\texorpdfstring{$_3$}{}}

We first recall the role of large diffeomorphisms in the AdS$_3$/CFT$_2$ correspondence. 

The starting point is AdS$_3$ in Poincar\'e coordinates: 
\begin{align}
    ds^2
    &=
        d\rho^2 + e^{2\rho/\ell_3} dw^+ dw^-
        \label{TTbarzero}
    \ ,
\end{align}
where $\ell_3$ is the AdS$_3$ scale.
The boundary CFT$_2$ on a cylinder inherits the light cone coordinates
$w^\pm = \phi\ell_3 \pm t$ from the bulk AdS$_3$. 
Interestingly, 
the baseline AdS$_3$ metric 
\eqref{TTbarzero} can be generalized to 
the Banados metrics \cite{Banados:1998gg}
\begin{align}
	ds^2
	&=
		d\rho^2 + \left(
			e^{\rho/\ell_3} dw^+
			+ \frac{\ell_3}{k}T_{--}(w^-) e^{-\rho/\ell_3} dw^-
		\right)
		\left(
			e^{\rho/\ell_3} dw^-
			+ \frac{\ell_3}{k} T_{++}(w^+) e^{-\rho/\ell_3} dw^+
		\right)
		\label{banados}
	\ ,
\end{align}
where $T_{\pm\pm}$ are general functions of the light cone coordinates $w^+$ and $w^-$, respectively. 
The curvature of the geometry is constant and negative for any functions
$T_{\pm\pm}(w^\pm)$, so the 
entire Banados family 
\eqref{banados} of metrics
describe AdS$_3$.
When $T_{\pm\pm}$ are both {\it constants}, the Banados metrics \eqref{banados} reduce to the BTZ black holes in AdS$_3$ \cite{Banados:1992gq,Banados:1992wn}. In this case the values of the constant $T_{\pm\pm}$
depend on physical mass $M$ and angular momentum $J$ or, equivalently, the coordinate positions $r_
\pm$ of the outer/inner horizons: 
\begin{equation}
\ell_3 T_{\pm\pm} \underset{\rm constant}{=}
 \frac{1}{2} (M\ell_3 \pm J) = \frac{k}{4\ell^2_3} (r_+ \pm r_-)^2 
~.\label{eqn:TpmBTZ}
\end{equation}

Indeed, any Banados metrics can be obtained from the $T_{++}=T_{--}=0$ baseline
metric \eqref{TTbarzero}
by a coordinate transformation. 
To see this, we consider reparametrizations of $\rho$ and $w^\pm$ that respect the gauge conditions $g_{\rho\rho}=1$ and $g_{\rho\pm}=0$. 
We also require that the metric's asymptotic behavior at large $\rho$ remains unchanged: $g_{+-} \sim \frac{1}{2}e^{2\rho/\ell_3}$. 
As a first attempt, one may simply consider $w^\pm\to f_\pm(w^\pm)$ for two functions $f_\pm$ satisfying appropriate periodicity and monotonicity conditions. This 
yields a conformal transformation of $g_{+-}$ that violates its asymptotic behavior at large $\rho$ and must be compensated by
$\rho\to\rho
		- \frac{\ell_3}{2}\ln\left(f_+'(w^+)f_-'(w^-)\right)$.
The resulting violation of the $g_{\rho\pm}=0$ gauge condition can be addressed by modifying the transformation of $w^\pm$, and this in turn forces a change in the $\rho$ reparametrization. Continuing this process to all orders identify
the exact diffeomorphisms
\begin{align}
\begin{split}
    w^\pm
    &\quad\to\quad
        f_\pm(w^\pm)
        - e^{-2\rho/\ell_3}
                \frac{
                    f_\pm'(w^\pm)^2f_\mp''(w^\mp)
                }{
                    2\ell_3^{-2}f_+'(w^+)f_-'(w^-)+ \frac{1}{2}e^{-2\rho/\ell_3}f_+''(w^+) f_-''(w^-)
                }
	\ ,\\
	\rho
	&\quad\to\quad
		\rho
		- \frac{\ell_3}{2}\ln\left(f_+'(w^+)f_-'(w^-)\right)
		+ \ell_3\ln\left(
			1+ \ell_3^2e^{-2\rho/\ell_3}\frac{f_+''(w^+) f_-''(w^-)}{4f_+'(w^+)f_-'(w^-)}
		\right)
	\ ,
	\label{3d_wrho}
\end{split}
\end{align}
where $f_+$ and $f_-$ are independent functions of $w^+$ and $w^-$, respectively.
They transform the line element
\eqref{TTbarzero} to
\begin{align}
	ds^2
	&\to
		d\rho^2
		+ e^{2\rho/\ell_3}
		\left(
			dw^+
			- \frac{\ell_3^2}{2}e^{-2\rho/\ell_3}\Sch(f_-,w^-) dw^-
		\right)
		\left(
			dw^-
			- \frac{\ell_3^2}{2}e^{-2\rho/\ell_3}\Sch(f_+,w^+) dw^+
		\right)
	\ ,
\end{align}
where $\Sch(\cdot,\cdot)$ denotes the Schwarzian derivative \eqref{eqn:Sch}.
This is precisely the Banados metric \eqref{banados} with $T_{\pm\pm}$ identified as
\begin{align}
	T_{\pm\pm}(w^\pm)
	&=
		-\frac{k\ell_3}{2}\Sch(f_\pm,w^\pm)
	\ .
 \label{eqn:Twm}
\end{align}
Any function $T_{\pm\pm}(w^\pm)$ that is regular at $w^\pm=0$ can be written as the Schwarzian derivative of some function $f_\pm(w^\pm)$. Therefore, 
the reparametrizations \eqref{3d_wrho} span all Banados metrics.

If, instead of the baseline Poincar\'e AdS$_3$ \eqref{TTbarzero}, the starting point is the general Banados metric \eqref{banados}, the diffeomorphism \eqref{3d_wrho} determines the transformation property
\begin{align}
T_{\pm\pm}(w^\pm) ~\to~ T_{\pm\pm}(w^\pm)f_\pm'(w^\pm)^2-\frac{k\ell_3}{2}\Sch(f_\pm,w^\pm)~.
\label{eqn:T3pm}
\end{align}
Thus $T_{\pm\pm}(w^\pm)$ can be identified with light-cone components of the CFT$_2$ energy-momentum tensor with the central charge
$$
k =\frac{1}{6} c~.
$$
Computations in gravitational variables establishes their normalization in terms of the gravitational coupling as \cite{Brown:1986nw}
\begin{equation}
k =\frac{1}{6} c = \frac{\ell_3}{4G_3}~.
\label{eqn:brownH}
\end{equation}
This detailed normalization is important in many applications of the AdS$_3$/CFT$_2$ correspondence but it is tangential for our study of AdS$_2$. 


For infinitesimal reparametrizations $f_\pm(w^\pm) = w^\pm + \ep_\pm(w^\pm)$, 
the transformations \eqref{3d_wrho} 
simplify to
\begin{align}
	w^\pm &\quad\to\quad w^\pm + \ep_\pm(w^\pm) - \frac{\ell_3^2}{2}e^{-2\rho/\ell_3} \ep_\mp''(w^\mp)
	\ ,\\
	\rho &\quad\to\quad \rho - \frac{\ell_3}{2}\ep_+(w^+) - \frac{\ell_3}{2}\ep_-(w^-)
	\ .
\end{align}
Thus the generating vector field $v$ has components $v^\pm = \ep_\pm(w^\pm)- \frac{\ell_3^2}{2}e^{-2\rho/\ell_3} \ep_\mp''(w^\mp)$ and $v^\rho = -\frac{\ell_3}{2}(\ep_+(w^+)+\ep_-(w^-))$.
It has norm-squared, computed using the baseline metric \eqref{TTbarzero}, that scales as $|v|^2 \sim e^{2\rho/\ell_3} \ep_+(w^+)\ep_-(w^-)$ for large $\rho$. This implies that the normalization condition on the vector field
diverges exponentially,
\begin{align}
	\int d^3x \sqrt g|v|^2 \ \sim\ 
		\frac{1}{2}\int dw^+ dw^- d\rho\, e^{4\rho/\ell_3} \ep_+(w^+)\ep_-(w^-)\ \to \ \infty~.
\end{align}
Therefore, the reparametrizations \eqref{3d_wrho} are generated by non-normalizable vector fields. That is what makes them ``large" diffeomorphisms.

On the other hand, the infinitesimal changes in the metric induced by the diffeomorphisms are $\delta g_{\pm \pm} = -\frac12 \ell_3^2 \ep_\pm'''(w^\pm)$.
They are finite, since $g^{+-} = 2e^{-2\rho/\ell_3}$ and the normalization condition converges,
\begin{align}
	\int d^3x \sqrt g \, |\delta g|^2
	\ =\ 
		\ell_3^4\int dw^+dw^-d\rho\, e^{-2\rho/\ell_3} \ep_+'''(w^+)\ep_-'''(w^-)
	\ <\  \infty
	\ .
	\label{3d_g_norm}
\end{align}
Thus the metric deformation remains within the allowable configuration space. Moreover, as discussed in the preceding paragraph, it is not pure gauge even though it is formally generated by a diffeomorphism. 

The Schwarzian derivative $\Sch(f_\pm,w^\pm)$ vanishes 
for the $\SL(2,\mathbb{R})_+ \times \SL(2,\mathbb{R})_-$ transformations
\begin{align}
f_\pm (w^\pm)\ell_3^{-1} = \frac{a(w^\pm\ell_3^{-1}) + b}{c(w^\pm\ell_3^{-1}) + d} ~,
\end{align}
where the coefficients $a$, $b$, $c$ and $d$ (with $ad-bc=1$) can be distinct for $w^+$ and $w^-$. The exact diffeomorphisms \eqref{3d_wrho} remain quite
non-trivial even in this case, but these apparent complications are just artifacts of the gauge. The
vector fields generated by $f_\pm$ do not change the metric \eqref{TTbarzero} at all, because (\ref{eqn:Twm}) vanish. If these diffeomorphisms had been normalizable they would be isometries. However, since they are not, they do not generate proper symmetries. They are outer automorphisms.

\subsection{Dimensional Reduction to AdS\texorpdfstring{$_2$}{}}

The goal of this subsection is to recast the large diffeomorphisms in AdS$_2$ discussed in earlier sections as a subset of their analogues AdS$_3$ that were reviewed in the previous subsection. 
Dimensional reduction from 3D to 2D was previously discussed by many authors, including \cite{Gupta:2008ki,Castro:2008ms,Balasubramanian:2009bg,Castro:2010vi,Castro:2014ima,Castro:2019vog,Cvetic:2016eiv,Ghosh:2019rcj}.

The linchpin to the dimensional reduction is rewriting of the Banados metric \eqref{banados} as
\begin{align}
\hspace{-2.5cm}
ds^2 & = d\rho^2  - \frac{k}{4\ell_3 T_{++}} \left( e^{2\rho/\ell_3} - \frac{\ell^2_3 T_{++}T_{--}}{k^2} e^{-2\rho/\ell_3}\right)^2 (dw^-)^2
\cr &\quad
+ \frac{\ell_3 T_{++}}{k}\left( dw^+ + \frac{k}{2\ell_3 T_{++}} \left( e^{2\rho/\ell_3} + \frac{\ell^2_3 T_{++}T_{--}}{k^2} e^{-2\rho/\ell_3}\right) dw^-\right)^2 ~.
\end{align}
We assume that  $T_{++}(w^+)$ is constant (independent of $w^+$), a necessary condition for compactification along $w^+$. Then comparison with the AdS$_2$ metric \eqref{ds2_SchTan} suggests that the AdS$_2$ time $\tau\sim w^-$ and the AdS$_3$ energy momentum tensor $T_{--}$ which, according to 
\eqref{eqn:Twm} is essentially the Schwarzian derivative, 
can be identified with its 2D analogue in 
\eqref{ds2_SchTan}. Our objective is to elevate these hints to a precise identification.  

However, it is not at all obvious that null compactification makes sense. As the notation suggests, the $w^\pm$ become light-cone coordinates at large $\rho$ in the AdS$_3$ setting, and they are exactly null in the limit $\rho\to\infty$. This corresponds to a spatial ``circle" with infinite radius, the exact opposite of what is needed to justify omission of ``heavy" KK-modes. In addition to the ``energetic" objection, higher modes in the Fourier expansion along $w^-$ are not negligible, null reduction suffers from a global discrepancy, at least in Lorentzian signature \cite{Balasubramanian:2009bg}: the AdS$_3$ coordinates $w^\pm$ both have period $2\pi\ell_3$, but the time $t_2$ in the AdS$_2$
spacetime is non-compact. 

These objections to compactification along a ``null" direction can be overcome because $w^+$ becomes {\it spatial} in the bulk of spacetime. This opening towards a justification develops to a proper argument {\it deeply} into the AdS$_3$ spacetime where the radius of the spatial $w^+$ is {\it small} relative to the prevailing length scale. The details of this procedure are subtle, and it is not clear that the community has reached a consensus on the subject. Therefore, we proceed carefully and provide details. The following discussion is adapted from \cite{Seiberg:1997ad,Balasubramanian:1997kd,Polchinski:1999br}.

We first insert factors of a real number $\lambda$ as follows: 
\begin{align}
\hspace{-0.8cm}
ds^2 &= d\rho^2  - \frac{k}{4J} \left( \lambda e^{2\rho/\ell_3} - \frac{\ell_3 J\lambda^2 T_{--}}{k^2} \frac{e^{-2\rho/\ell_3}}{\lambda}\right)^2 \left(\frac{dw^-}{\lambda}\right)^2
\cr&\quad
+ \frac{J}{k}\left( dw^+ + \frac{k}{2J} \left( \lambda e^{2\rho/\ell_3} + \frac{\ell_3 J\lambda^2 T_{--}}{k^2} \frac{e^{-2\rho/\ell_3}}{\lambda} \right) 
 \frac{dw^-}{\lambda}\right)^2 ~.
\nonumber
\end{align}
This step is completely trivial, the metric remains the same for any value of $\lambda$. With some prescience, and 
to lighten the formulae a bit, we denoted the constant $T_{++} = J\ell^{-1}_3$, but we could have retained $T_{++}$ and deferred this simplification to later. 
Next, we remove the $\lambda$'s by introducing new notation
\begin{align}
\begin{split}
t_2 &= \lambda^{-1} w^- ~,\cr
e^{2\rho_{\rm new}/\ell_3} & =\lambda \sqrt{\frac{k}{J}} e^{2\rho/\ell_3}~,\cr
T_2(t_2) &= \lambda^2 \ell_3 T_{--}~,\cr
\ell_2 & = \frac{1}{2} \ell_3~.
\end{split}
\end{align}
The substitutions give
\begin{align}
ds^2 = d\rho^2  - \frac{1}{4} \left( e^{\rho/\ell_2} - \frac{  T_2(t_2)}{k} e^{-\rho/\ell_2}\right)^2 dt_2^2
+ \left( \sqrt{\frac{J}{k}} dw^+ + \frac{1}{2} \left( e^{\rho/\ell_2} + \frac{T_2(t_2)}{k} e^{-\rho/\ell_2} \right) 
 dt_2  \right)^2 .
 \nonumber
\end{align}
This second step is also just preparatory, a change of notation that does not modify the geometry at all.

At this point, we take the limit $\lambda\to 0$
with the ``old" coordinates fixed. This may appear trivial, but it is very significant. 
For example, the original $w^-$ was periodic with period $2\pi\ell_3$ so $t_2$ acquires period $2\pi\ell_3 \lambda^{-1}$ and, in the limit $\lambda\to 0$, it is not periodic at all, it has been decompactified. This was a major goal of the entire maneuver. 

With the ``old" $T_{--}$ fixed, the ``new" $T_{--}$, which is actually denoted $T_2$, becomes small and $\ell_3 T_{--}\to 0$ 
is the extremal limit, according to \eqref{eqn:TpmBTZ}.
Therefore, $\ell_3 T_{++}\underset{\rm extremal}{=}J$ as anticipated already. The remaining $T_2$, the 2D image of $T_{--}$, corresponds to excitations over this ground state. Since the periodicity of $w^+$ is independent of $\lambda$, the lower typical energy after the scaling limit justifies ``null" compactification. 

The final result for the AdS$_2$-geometry after compactification of the Banados metric \eqref{banados} with constant $T_{++}$ is:
\begin{align}
ds^2 = d\rho^2  - \frac{1}{4} \left( e^{\rho/\ell_2} - \frac{T_2(t_2)}{k} e^{-\rho/\ell_2}\right)^2 dt_2^2.
\end{align}
This can be identified with the AdS$_2$ geometry \eqref{eqn:Fmetrc}, in units where $\ell_2=1$ with the identification $t_2\equiv \tau$. 
This is the central result that allow us to identify the large diffeomorphisms in AdS$_2$ \eqref{2d_finite_real} with
their analogues in AdS$_3$ \eqref{3d_wrho}.  

In our discussion of the limits, we took the view that $t_2$ is very large, corresponding to very low energy $T_2$, but it is equally valid to change perspective and interpret $t_2$ and $T_2$ as ``typical", while all the original energies are very large. From either point of view, 
the ``motion" starting from $\lambda=1$ towards smaller $\lambda$ prioritizes the near horizon region where $\rho$ is smaller, times $t_2$ that are longer, and energies $T_2$ that are smaller. This is possible because the black hole spacetime is tuned simultaneously.  

The dimensional reduction shows that the AdS$_2$ spacetime is supported by a gauge field that depends on spacetime position
\begin{align}
A= \left(e^{\rho/\ell_2} + \frac{T_2(t_2)}{k} e^{-\rho/\ell_2}\right)dt_2~.
\nonumber
\end{align}
The corresponding field strength is proportional to the 2D volume form so this gauge field acts like a 2D cosmological constant. 

The dimensional reduction from AdS$_3$ to AdS$_2$ relates the Brown-Henneaux value for the level $k$ 
\eqref{eqn:brownH} in AdS$_3$ to the gravitational coupling constant in AdS$_2$:
\begin{equation}
S_0 = \frac{1}{4G_2} = \frac{2\pi R_3}{4 G_3}  = 2\pi \sqrt{Jk}~.
\label{eqnn:2sS0} 
\end{equation} 
This is the correct value for the inert ground state entropy. 

The starting point of the discussion in this section was the Poincar\'e vacuum of AdS$_3$ \eqref{TTbarzero}, and its excitations in the form of the Banados metrics \eqref{banados}. Asymptotically, these geometries approach a cylinder, so the implicit understanding has been that the AdS$_3$ energy momentum tensors $T_{\pm\pm}$ are defined on the cylinder. With this convention the true AdS$_3$ vacuum, the one that has the lowest energy and is invariant under $\SL(2)^2$, assigns the negative eigenvalue $-\frac{1}{4}k$ to both of $\ell_3T_{\pm\pm}$. The dimensional reduction from AdS$_3$ to AdS$_2$ requires setting $T_{++}$ to a positive value that defines the filled states with that chirality. The default is to take $\ell_3T_{--}$ positive as well, corresponding to excitations above the R-vacuum with $\ell_3T_{--}=0$, but it is also possible to consider the NS-vacuum $\ell_3T_{--} = - \frac{k}{4}$. After adjusting normalizations of coordinates, 
this option amounts to the special case where AdS$_3$ takes the form
$$
ds^2 = d\rho^2  - \frac{1}{4}  \left( e^{2\rho/\ell_3} + e^{-2\rho/\ell_3}\right)^2 (dw^-)^2
+ \left( dw^+ + \frac{1}{2} \left( e^{2\rho/\ell_3}  - e^{-2\rho/\ell_3}\right) dw^-\right)^2 ~, 
$$
with $w^-$ noncompact and $w^+$ having period $2\pi\ell_3$. This is the self-dual orbifold \cite{Coussaert:1994tu}: it is a coset of AdS$_3$ such that the self-dual $\SL(2)$ isometries (those acting on $w^-$) are preserved.
In our discussion we depart from the proposal of \cite{Balasubramanian:2009bg} and instead identify the self-dual orbifold with the AdS$_2$ wormhole
discussed in subsection \ref{subsec:wormhole}.

\acknowledgments

We thank participants in the Princeton University (PCTS) workshop on ``Low Dimensional Holography and Black Holes" for comments and discussions in response to a version of this article presented by FL. 
FL thanks the Simons Foundation for support through a sabbatical fellowship. He also thanks Stanford Institute for Theoretical Physics for hospitality and support in the course of the sabbatical. 
The work of SC is supported by the European Research Council (ERC) under the European Union’s Horizon 2020 research and innovation programme (grant agreement No 852386).
SC also acknowledges financial support from the Samsung Scholarship.
This work was supported in part by DoE grant DE-SC0007859.

\bibliographystyle{JHEP}
\bibliography{references}

\end{document}